\documentstyle[aps,epsf]{revtex}
\begin{document}

\title{ Symmetries and 
Decoherence of Two-Component Confined Atomic Clouds:
\\
 Study
of the Atomic Echo in the two-component Bose-Einstein Condensate. }
\author{A. B. Kuklov$^1$, N. Chensincki$^1$
 and Joseph L. Birman $^2$}
\address{$^1$ Department of Engineering 
Science and Physics,
The College of  Staten Island, CUNY,
     Staten Island, NY 10314}
\address{$^2$ Department of Physics, City College, CUNY, New York, NY 10031}

\maketitle

\begin{abstract}
Implications of the internal symmetries on the
dynamics of the trapped two-component atomic vapors
are discussed. In the cases of $^{87}$Rb (bosons) 
as well as of $^{40}$K (fermions)
trapped in the two hyperfine states, the intrinsic
$su(2)$ symmetry can be realized with a very good precision.
Such a symmetry protects the global operators, which are
the generators of the symmetry, 
from any decoherence.
The case of boson-fermion mixture 
is discussed as well.
The role of external factors, breaking the symmetry,
in inducing the decoherence of the global operators
is considered.
It is shown that, 
the loss of the correlations is not faster 
than the rates of the induced heating or losses, 
provided
the noise is characterized by short
correlation length. 
The case of the extrinsic long-ranged fluctuations
is also considered.
Intrinsic mechanisms of decoherence of the 
correlators of the condensate operators
of the two-component condensate are analyzed. 
The atomic echo
is discussed as a test for the reversibility
of the phase diffusion effect. The intensity and the
profile of the multiple echo are calculated 
numerically as well as analytically.


\end{abstract}\vskip0.5 cm 

\section{Introduction}

Successful achievements of confinement
and cooling to the state of degeneracy 
of Bose \cite{BEC} and Fermi \cite{FERMI} 
atomic gases have created unprecedented 
variety of possibilities for studies
of the many-body effects. The fundamental
issue of emergence of irreversibility
can now fully be addressed experimentally
and theoretically in the wide range of
situations. 

Studies of the dissipation of the
normal modes of the confined Bose-Einstein
condensates (BEC) \cite{DAMP} began 
immediately after BEC has been realized.
Very recently, the data on the damping
in the Fermi mixtures has been reported
as well \cite{JIN}. 

The Ramsey spectroscopy study of the temporal 
correlations of the relative phase of the two-component
~$^{87}$Rb BEC has been conducted by
the JILA group \cite{JILA}.
No
decay of the global relative 
phase has been detected on the time scale
of the experiment $\leq 100$ms. 
The question was then posed in \cite{JILA} 
as to why the phase correlations are
so robust despite an apparent fast relaxation of
other degrees of freedom such as, e.g.,
the relative motion of the two condensates.

Many intriguing questions
are associated with the finiteness of
the number of bosons $N$ forming 
BEC in the trap.
It has been realized that the global phase
may experience collapses and revivals as long as
$N$ is finite \cite{PD}. 

Symmetries play a very special role in the
many-body dynamics.
As was discovered by Pitaevskii and Rosch
\cite{PIT1}, a 2D gas trapped by the oscillator
potential and characterized by the contact
two-body interaction can be described by
the dynamical symmetry ~$SO(2,1)$, and this
results in the non-decaying
breathing mode. The symmetry
{\it protects} the mode from the decay.   
In our recent work \cite{SU2}, it has been 
found that, in the case of the
$^{87}$Rb mixture characterized by the 
proximity to 
the intrinsic ~$su(2)$~ symmetry \cite{$su(2)$}, the dynamics of
the global generators is also protected by
this symmetry from the intrinsic 
decoherence. In the case
of the broken symmetry, the
rate of the decoherence is controlled by the
deviations from the ~$su(2)$, and, accordingly,
the decoherence is slow as long as the deviations
are small. This answers
the question \cite{JILA} about the robustness
of the phase dynamics. 

In another our work \cite{OC}, the nature
of relaxation of the condensate operators
was discussed. I has been shown that,
if the number of bosons ~$N_0$~ in the 3D 
equilibrium condensate is macroscopic, the
decoherence of the condensate time-correlators  
is rather a reversible dephasing than a
true irreversible decay. In some sense,
the Bose permutation symmetry,
representing the indistinguishability of particles
which populate the lowest (condensate) single
particle state, protects the condensate
operators from the irreversible loss of memory.
It was suggested
to test this by the echo-type experiment.
Thus, even though the intrinsic ~$su(2)$~ symmetry
can be broken, the dynamics of the condensate
operators remains reversible.
 
In the present work, we, first, extend
our analysis \cite{SU2} to the case of 
the two-component fermion mixture \cite{JIN},
and will show that, if the trapping potentials
for the components are identical, the global
quantities - the generators of the symmetry formed
by the Fermi operators - do not decay in spite of
the presence of the interaction between the
components. We also analyze the case of 
the boson-fermion mixture, which is close
to the ~$su(2)$~ symmetry with respect to the 
mutual (formal)  transformation of bosons and fermions
into each other.
In this situation, the decoherence
of certain global operators, which, however,
are not the generators, is suppressed as well.

Second, we analyze the problem of the
decoherence 
in the case of deviations from the 
~$su(2)$~ symmetry in the two-component
BEC. Extrinsic and intrinsic mechanisms 
of the decoherence are considered. 
For the extrinsic
factors, we concentrate on the interaction
with the hot background gas, which induces heating
and losses (or deposition).
The heating is modeled 
by introducing a random white noise
potential, which is characterized by
some spatial correlation length ~$L_p$. 
We show that the ratio of the 
phase decay rate ~$\tau^{-1}_{irr}$~ to
the heating rate ~$\tau^{-1}_h$~
is essentially given by the factor ~$L_p^2/L_c^2$,
where ~$L_c$~ stands for the thermal length of
the trapped atoms at the BEC transition
 temperature ~$T_c$. Obviously, ~$L_p$~ is given
by the thermal length of the background gas,
which is ~$\ll L_c$.
This implies that 
the phase memory loss turns out to be
{\it not faster}
than the heating rate. 
Similarly, incoherent
losses from the trap are unable to produce
the phase decay faster than the rate of the losses.

We also analyze
the role of the instrumental effects, which 
may produce the long range noise breaking the
intrinsic symmetry. In this situation, the phase
decay rate can be significantly enhanced, if compared
with the corresponding rate of disrupting 
the thermal equilibrium within the atomic cloud.
 However, we estimate that 
the phase decay rate is still too small 
in the current traps, provided the
instrumental effects are on the level of the
equilibrium room temperature thermal noise. 

Regarding the intrinsic factors,
we extend our analysis \cite{OC} to the
case of the two-component BEC, and show that
the intrinsic factors {\it do not} induce
the irreversible decay of the 
correlators of the condensate operators.
At finite
temperatures ~$T$, the phase exhibits 
{\it reversible} phase diffusion effect
similar to that analyzed previously at ~$T=0$ 
by many authors \cite{PD}.
In fact, in any exact many-body eigenstate,
the evolution of the condensate operators
is given by the relative chemical potential. Averaging
over the thermal ensemble produces
dephasing, whose rate is determined
by the fluctuations of the chemical potential.
At $T=0$, these fluctuations are produced by
the averaging over the initial {\it coherent}
state and/or shot noise, 
and, at ~$T\neq 0$, the thermal averaging
induces an additional contribution. 
Thus, at finite ~$T$~ the nature of the
global relative phase diffusion remains, practically, the same
as that at ~$T=0$, provided intrinsic factors only 
contribute to the decoherence.

Finally, we analyze the atomic echo
effect, suggested previously for testing the reversibility
of the phase diffusion \cite{OC,SU2}, 
in the context of the JILA experiment
\cite{JILA}. 
The strength of the echo is evaluated.
When the dephasing is
dominated by the shot noise in the total number
of bosons in the trap, the echo can be as large as
100\%, for the ~$\pi$-pulse employed for the 
time reversal \cite{SU2}. When the phase diffusion is dominated
by the self-interaction effects at ~$T=0$,
the echo about 50\% can be
achieved by a weak
time-reversal pulse,
which transfers coherently a (relatively) small
number of bosons between the components.
An analysis of the echo at finite
temperature is presented as well.
Our conclusion is that the 
echo, although suppressed, can still
be observed
at finite temperatures, 
provided the population
of the condensate state is dominant.

\section{Intrinsic symmetries and the dynamics
of their generators in the two-component atomic
mixtures}

Here we consider simplest cases of the
two-component mixtures: bosons of sort 1 + bosons
of sort 2;
fermions of sort 1  + fermions of sort 2; and one component
bosons + one component fermions. 

\subsection{Two-component Bose gas}

First, for the purpose
of consistency, we will briefly outline
the results \cite{SU2} for the  
two-component Bose mixture.
We employ the following two-component
Hamiltonian

\begin{eqnarray}
\displaystyle
H&=&\int\, d{\bf x}\{\Psi^{\dagger}_1(H_1 -{\epsilon_0\over 2})
 \Psi_1
+\Psi^{\dagger}_2(H_2+{\epsilon_0\over 2}) \Psi_2
+({\hbar\Omega^* (t)\over 2}\Psi^{\dagger}_2\Psi_1 + H.c)+
\nonumber
\\
\phantom{XXXX}
\label{0}
\\
&+&{2\pi \hbar^2 \over m}[a_1\Psi^{\dagger}_1\Psi^{\dagger}_1\Psi_1
\Psi_1
+a_2\Psi^{\dagger}_2\Psi^{\dagger}_2\Psi_2\Psi_2 +2a_{12}
\Psi^{\dagger}_1\Psi^{\dagger}_2\Psi_2\Psi_1]\},
\nonumber
\end{eqnarray}
\noindent
where 
~$ \Psi_{i}$~ is the second quantized
Bose field of the $i$-the component ($i=1,2$);
~$H_j=-\hbar^2\nabla^2/m_j \,\, + U_j({\bf x})$~ 
stands for the one-particle part which
includes the kinetic energy and the trapping
potential; the quantity ~$\epsilon_0=const$~
is the detuning of the external field ~$\hbar\Omega (t)$~
(taken in the rotating wave approximation ) from the
resonance between the components;
~$ \Omega(t)$~ stands for the 
corresponding Rabi frequency
treated as an envelope of the rf-pulses;
The binary collision terms
in (\ref{0}) are taken in the contact form, with ~$a_1,\,a_2,\,
a_{12}$~ being the corresponding scattering lengths.

The intrinsic $su(2)$ symmetry corresponds
to the situation ~$H_1=H_2,\,\, a_1=a_2=a_{12}$.
The generators of this symmetry 

\begin{eqnarray}
I_z={1\over 2}\int d{\bf x}(\Psi^{\dagger}_2\Psi_2-
\Psi^{\dagger}_1\Psi_1);\,\,\,
I_+=\int d{\bf x}\Psi^{\dagger}_2\Psi_1 ; \,\,\,
I_-=\int d{\bf x}\Psi^{\dagger}_1\Psi_2 \,
\label{3}
\end{eqnarray}
\noindent
represent the $su(2)$ algebra
of the angular momentum operators.
Indeed, it is easy to see that these operators obey the
standard $su(2)$ commutation relations 

\begin{eqnarray}
[I_z,I_+]=I_+,\,\,\,\,
[I_z,I_-]=-I_-,\,\,\,\, [I_+,I_-]=2I_z,
\label{3_1}
\end{eqnarray}
\noindent
provided the 
field operators obey the Bose commutation rule
~$[\Psi_i({\bf x}),\Psi^{\dagger}_j({\bf x}')]=\delta_{ij}
\delta ({\bf x}-{\bf x}')$. 
The operator ~$I_z=(N_2-N_1)/2 $,
with ~$N_{1,2}$~ standing for the total number
of bosons in the first and the second components,
respectively,
can be viewed
as the z-component of the angular momentum operator;
~$I_x=(I_++I_-)/2,\,\,I_y=(I_+-I_-)/2i$~
are the x,y-components, respectively.

If the condensate is present, the operators
~$I_{\pm}$~ carry the information
about the relative global phase ~$\varphi$~
(we may call it simply "phase"). Indeed,
in this case, ~$\Psi_j \approx \sqrt{N_j/V}
{\rm e}^{i\varphi_j}$, where ~$\varphi_j$~
is the global phase of the $j$th component, and
~$V$~ stands for the effective volume 
occupied by the condensate.
Thus, ~$I_+\approx \sqrt{N_1N_2}{\rm e}^{i\varphi},
\,\,\, \varphi = \varphi_1 -\varphi_2 $. 
Accordingly, the correlator ~$
\langle I_+(t)I_-(0) \rangle $, where the
mean is taken with respect to some set of initial
states, carries the information about
the time correlation properties of the
phase.   

In the case of the exact intrinsic
symmetry, the generators (\ref{3})
obey linear Heisenberg equations 

\begin{eqnarray}
i\hbar \dot{I}_z={\hbar \Omega^* (t) \over 2}I_+ -
{\hbar \Omega (t) \over 2}I_-,\,\,\,
\,\, i\hbar \dot{I}_+=-\epsilon_0I_+
+\hbar\Omega(t)I_z,\,\,\,
i\hbar \dot{I}_-=\epsilon_0I_-
-\hbar\Omega^*(t)I_z\, ,
\label{4}
\end{eqnarray}
\noindent
where the $su(2)$ commutation relations (\ref{3_1})
have been
employed \cite{SU2}. If ~$\Omega (t)$~ and ~$\epsilon_0$~
are not subjected to any noise,
the solution for the correlator ~$\langle I_+(t)I_-(0)\rangle$~
exhibits no decoherence despite the presence of the two-body
interaction in the Hamiltonian (\ref{0}).
It should also be noted
that Eq.(\ref{4}) are exact, and
they do not require a presence 
of the BEC. 

The decoherence will arise,
if either the one-particle parts ~$H_{1,2}$~
are different or the scattering lengths
deviate from the condition
~$a_1=a_2=a_{12}$. The nature of the
decoherence in the case of the BEC
will be discussed later.

\subsection{Two-component Fermi mixture}

In the case of the two-component Fermi
gas, the $su(2)$ intrinsic symmetry can
hold as well. For example, trapping
$^{40}$K \cite{FERMI} in the optical trap
results in the identical potentials
for both components \cite{JINPRIV}.
This, in fact, implies that the
intrinsic $su(2)$ symmetry holds
for any value of the inter-component
scattering length. Indeed, 
in the Hamiltonian (\ref{0}),
the Bose operators ~$\Psi_{1,2}$~ are to be replaced by 
the Fermi operators ~$F_{1,2}$~ obeying the anti-commutation
rule

\begin{eqnarray}
\{F_i({\bf x}), F^{\dagger}_i({\bf x'})\}=
\delta_{ij}\delta ({\bf x}- {\bf x'}),
\label{5}
\end{eqnarray}
\noindent
where ~$\{ ... , ...\}$~ denotes the anti-commutator.
Accordingly, the terms ~$\sim a_1, \,\, a_2$~ vanish
in Eq.(\ref{0}).
Introducing the $su(2)$ generators (\ref{3}),
where the Bose-operators are replaced by
the Fermi-operators, the ~$su(2)$~ commutation
relations (\ref{3_1}) follow. Finally,
the equations of motion for the generators
turn out to be Eq.(\ref{4}). Thus, in the case
of the two-component fermion gas, the dynamics
of the generators is always {\it linear} regardless
of the value of the scattering length ~$a_{12}$,
as long as the one-particle Hamiltonians ~$H_{1,2}$~  are
the same. We should, however, note that
the higher order scattering harmonics change this situation,
and the corresponding restrictions should be imposed on 
the scattering amplitudes of the higher orders (p-,f- waves etc).
These processes, however, are insignificant at low
temperatures and at low densities.

\subsection{Bose-Fermi mixture}

In the case of the mixture of, e.g., one-component
Bose and Fermi gases, the respective fields
~$\Psi ({\bf x}), \,\,\, F( {\bf x})$~ 
are introduced, and the Hamiltonian takes the form

\begin{eqnarray}
\displaystyle
H&=&\int\, d{\bf x}\{\Psi^{\dagger}(H_1 -{\epsilon_0\over 2})
 \Psi
+F^{\dagger}(H_2+{\epsilon_0\over 2})F   
+ {2\pi \hbar^2 \over m}[a_1\Psi^{\dagger}
\Psi^{\dagger}\Psi\Psi
+2a_{12}
\Psi^{\dagger} F^{\dagger}F\Psi]\},
\label{00}
\end{eqnarray}
\noindent
where it was taken into account that
no interconversion between bosons and fermions is possible,
and that the self-interaction between the one-component
fermions vanishes in the s-wave approximation.

Let us analyze a special situation -- ~$H_1=H_2,\,\,\,
a_1=a_{12}$. This corresponds to the (accidental)
intrinsic ~$su(2)$~ symmetry between the bosons and
the fermions. Indeed, the interaction term
can be rewritten as ~$a_1(\Psi^{\dagger}\Psi)^2
+2a_{12}
\Psi^{\dagger}\Psi F^{\dagger}F = a_1[\Psi^{\dagger}\Psi
+
F^{\dagger}F]^2$~ in this case, because ~$[F^{\dagger}F]^2$~
vanishes (apart from a trivial term ~$\sim F^{\dagger}F$,
which simply renormalizes the constant ~$\epsilon_0$~
in Eq.(\ref{00})). Now, if ~$H_1=H_2$, the formal
$su(2)$ transformation acting on the "spinor"
made of ~$\Psi,\,\, F$~ retains the Hamiltonian 
intact. 

We introduce the operators

\begin{eqnarray}
I_z={1\over 2}\int d{\bf x}(\Psi^{\dagger}\Psi-
F^{\dagger}F),\,\,\,
I_+=\int d{\bf x}\Psi^{\dagger}F , \,\,\,
I_-=\int d{\bf x}F^{\dagger}\Psi,\,\, \,
I= \int d{\bf x}( \Psi^{\dagger}\Psi +
F^{\dagger}F)
.
\label{300}
\end{eqnarray}
\noindent
It should be noted that these operators
{\it do not} form a closed algebra, as
can easily be verified by commuting them with each
other. Instead,
they obey the following relations

\begin{eqnarray}
[I_z,I_+]=I_+,\,\,\,\,
[I_z,I_-]=-I_-,\,\,\,\, \{I_+,I_-\}= I,\,\,
[I_z, I]=[I_{\pm},I]=0.
\label{3_10}
\end{eqnarray}
\noindent

It can be verified that the Heisenberg
equations of motion for the above operators
are also {\it linear}. Specifically,

\begin{eqnarray}
\dot{I}_z= 0,\,\,\, \dot{I}=0, \,\,\,
 i\hbar \dot{I}_{\pm}= {\mp}\epsilon_0I_{\pm},
\label{40}
\end{eqnarray}
\noindent
as long as the above condition of the intrinsic
~$su(2)$~ symmetry holds. 
Thus, the situation of the Bose-Fermi mixtures
can also exhibit the dissipationless dynamics
of certain global quantities (which are not,
however, the generators of the symmetry). 
It should be mentioned that any deviations
from the intrinsic symmetry will result in the
decoherence of these global quantities, and
the rate of such a decoherence will also be
controlled by the proximity to the symmetry.

\section{The Nature of Decoherence in the
Case of the Broken $su(2)$ Symmetry in the
Two-Component Bose-mixture}  

Here we will discuss various factors which
break the intrinsic ~$su(2)$, and thereby
introduce the decoherence in, e.g., the
correlator 

\begin{eqnarray}
\rho_{ij}({\bf x}, {\bf x}', t,t')
 =\langle \Psi^{\dagger}_i({\bf x}, t)
\Psi_j({\bf x}', t')\rangle, \quad t>t'.
\label{D1}
\end{eqnarray}
\noindent
Hereafter we consider the
Bose statistics only.

First, we address
the decay induced by extrinsic
factors such as: 1) collisions with hot background
gas, which result in heating of the
confined cloud and in losses from the trap;
2) Thermal noise of the trapping potential,
which corresponds to temperatures much higher
than the confined gas.

We raise a general question: {\it Under what
conditions can the loss of the memory, 
produced by the extrinsic
factors, be much faster than the corresponding
time of disrupting the equilibrium in BEC}?
In other words, if ~$\tau_h$~ is the
time of the heating induced by some extrinsic
factor,
and ~$\tau_{irr}$~ stands for the time
of the irreversible loss of the time
correlations in (\ref{D1})
due to the same factor,
can it be that the condition 

\begin{eqnarray}
\tau_{irr} \ll \tau_h
\label{D2}
\end{eqnarray}
\noindent
 holds?
In the same sense, if ~$\tau_L$~ stands
for a typical life-time of the
confined cloud, which is being subjected
to losses, 
can the condition (\ref{D2}) 
hold, where the role of ~$\tau_h$~ is replaced
by ~$\tau_L$~? 

Below we will show that the range of the spatial
correlations of the extrinsic factors plays a
crucial role in fulfilling the condition (\ref{D2}).
In fact, the case 2) turns out to be the most
efficient in destroying the time correlations
without introducing significant disturbances
into the system.

\subsection{Rates of disrupting the equilibrium,
and the loss of the phase memory}

The purpose of the following
is to outline the framework within
which the effects of environment
on the confined cloud can be analyzed,
and to obtain general criterion allowing
to compare the rates of disrupting
the equilibrium and of the decoherence.
In this section, we will consider
the effects of interaction 
with the background (hot) gas\footnote{ We
believe that the conclusion ensuing 
from this analysis is general}.
In order to do this consistently,
the corresponding interaction term
should be added to the Hamiltonian (\ref{0}).
We choose it in the form

\begin{eqnarray}
\displaystyle
H_L=\sum_{ij, kl}\int\, d{\bf x}g_{ij,kl}\Psi^{\dagger}_i
B_k^{\dagger}B_l \Psi_j ,
\label{D13}
\end{eqnarray}
\noindent
where the summation runs over all the components
of the confined cloud and of the background
gas described by the fields ~$B_i$; ~$g_{ij,kl}$~
are the corresponding interaction constants. 

The term (\ref{D13}) describes two basic effects \cite{CORN}:
a) a fast particle strikes a confined boson and
ejects it from the trap; b) the background particle
does not eject any boson from the cloud
and, instead, imparts a substantial
energy producing heating in the trap,
and, possibly, exchange between the components. 
The term (\ref{D13}) can be employed in order
to derive the equation for the 
one-particle density matrix
(OPDM) ~$\rho_{ij}({\bf x}, {\bf x}', t)
 =\langle \Psi^{\dagger}_i({\bf x}, t)
\Psi_j({\bf x}', t)\rangle=\rho_{ij}({\bf x}, {\bf x}', t,t)
$~ as well as for the correlator
(\ref{D1}). The OPDM
equation becomes a set of the kinetic equations
in the Wigner representation.
 Here we will not implement this
to full extent. Instead, 
we will analyze an exactly solvable
model which captures basic features
of the effect b), and then we will 
present results for the effect a).

The field ~$\xi_{ij}=
g_{ij,kl}B^{\dagger}_kB_l$~ in Eq.(\ref{D13}) can be viewed 
as some random mean field potential 
acting on the confined cloud.
As long as the background gas is hot,
this effective random potential can be
considered as a white noise. As it will be
seen later, this assumption allows
obtaining the OPDM equation exactly
in the case when collisions between the confined
bosons are ignored.

In order to simplify further analysis,
we ignore the exchange of the components
induced by the collisions\footnote{
while complicating the calculations,
inclusion of the random exchange
 does not change the main conclusion}, 
and consider
just  a one-component random potential
~$\xi$~ which satisfies the condition 

\begin{eqnarray}
\langle \xi({\bf x}, t)\xi({\bf x}', t')\rangle
=\delta (t-t')C(|{\bf x}-{\bf x}'|),
\label{D3}
\end{eqnarray}
\noindent
where ~$ C(|{\bf x}-{\bf x}'|)$~ stands for some
smooth function which decays to zero on 
some typical length ~$L_p$. This function
describes spatial correlations of the noise. 
 
It
should be noted that the above model
{\it is not} suitable for treating a thermal
equilibrium between the background and the cloud.
Indeed, the external time dependent potential
may produce unlimited heating, similarly to the case
of the Brownian motion, if no friction term
counterbalancing the external noise is introduced.
In order to treat the equilibrium, a fully Hamiltonian
analysis based on Eq.(\ref{D13}) 
should be implemented.
This, however, is not required because the
confined cloud is essentially out of the equilibrium
with the background, and we are estimating the rate
of disrupting 
the equilibrium in the
cloud.

The Hamiltonian is taken in the form (\ref{0}), where 
the noisy potential should be added. We
choose ~$a_1=a_2=a_{12}=0$. Furthermore, for the sake
of simplicity, we consider no trapping potential
and set without the loss
of generality ~$\epsilon_0=0,\,\, \Omega =0$~, so that
the Hamiltonian takes a form

\begin{eqnarray}
\displaystyle
H=\int\, d{\bf x}\{\Psi^{\dagger}_1H_0 
 \Psi_1
+\Psi^{\dagger}_2H_0 \Psi_2
+\xi_1\Psi^{\dagger}_1\Psi_1 + \xi_2
 \Psi^{\dagger}_2\Psi_2] \}, \quad \xi_1=-\xi_2=\xi,
\label{D4}
\end{eqnarray}
\noindent
where ~$H_0=-{\hbar^2 \over 2m}\nabla^2$, and we have included
the interaction with the random potential ~$\xi$~ through the
term breaking the intrinsic ~$su(2)$~ symmetry, because,
according to the preceding analysis, no decoherence
of the global operators can be induced
in the case of the intrinsic symmetry (~$\xi_1=\xi_2$~ in the
considered situation). 

In the presence of the noise, the meaning
of the averaging in Eq.(\ref{D1}) should
be specified. Let us assume that some
initial state ~$|t=0\rangle$~ was created
at the time moment ~$t=0$, and the noise
affects the following evolution at later times
~$t>0$. Thus, the averaging should be performed, first, over
the initial state (or a set of states with
some weight), and, then, the averaging over
the gaussian noise is to be done. 

The OPDM
 characterizes the rate of the heating. Indeed,
the mean kinetic energy of the particles 
is defined as

\begin{eqnarray}
\displaystyle
K=\sum_{i=1,2}\int\, d{\bf x}H_{0{\bf x}'} \rho_{ii}({\bf x}, 
{\bf x}', t), \quad {\bf x}'\to {\bf x},
\label{D5}
\end{eqnarray}
\noindent
where, the coordinate dependence ~$H_{0{\bf x}'}$~
in the kinetic energy operator ~$H_0$~
indicates that this operator acts
on the coordinate ~${\bf x}'$, and, then,  one should set
~${\bf x}'={\bf x}$~ and integrate over ~${\bf x}$. 

Equations for the OPDM follow from the Heisenberg
equations ~$i\hbar \dot{\Psi}_a=(H_0+\xi_a)\Psi_a$. 
Employing the Furutzu-Novikov
theorem \cite{NOV} (see details in
Appendix A), 
we obtain the exact equations for the OPDM as

\begin{eqnarray}
\displaystyle
i\hbar\dot{\rho}_{ii}({\bf x},{\bf x}', t)&=&
[-H_{0{\bf x}} + H_{0{\bf x}'}] \rho_{ii}
({\bf x},{\bf x}', t) - {2i \over \hbar}
[C(0) - C(|{\bf x}-{\bf x}'|)]
\rho_{ii}({\bf x},{\bf x}', t);
\label{D61}
\\
\phantom{XXX}
\nonumber
\\
i\hbar \dot{\rho}_{12}({\bf x},{\bf x}', t)&=&
[-H_{0{\bf x}} + H_{0{\bf x}'}] \rho_{12}
({\bf x},{\bf x}', t) - {2i \over \hbar} 
[C(0) + C(|{\bf x}-{\bf x}'|)]
\rho_{12}({\bf x},{\bf x}', t).
\label{D63}
\end{eqnarray}
\noindent
It should be noted that reinstating the
inter-particle interaction will result
in a chain of coupled equations for the
correlators, where 
the mean involves averaging over the noise.
On the each next step, the decoupling of the noise
term can be done by means of employing
the Furutzu-Novikov theorem \cite{NOV}.
This will lead to finding the corrections to
the collision integral due to the noise. 
Practically, it is enough to neglect
such corrections.
Accordingly, Eqs.(\ref{D61}), (\ref{D63})
can be modified by simply adding the collision
integrals obtained with no noise present.    
Inclusion of such integrals, however, does not affect
the main result.

We choose the initial state as a uniform
condensate characterized by finite
correlations between the components. Accordingly
~$\rho_{ij}
({\bf x},{\bf x}', t=0)=\rho^{(0)}_{ij}=const
$~ for all the components. 
As time proceeds, this state becomes destroyed.
Indeed, the solution of Eqs.(\ref{D61}), (\ref{D63})
corresponding to the chosen initial condition
is

\begin{eqnarray}
\displaystyle
\rho_{ii}({\bf x},{\bf x}', t)=
{\rm e}^{- 2[C(0) - C(|{\bf x}-{\bf x}'|)]t/ 
\hbar^2 }\rho^{(0)}_{ii},
\label{D71}
\\ \\
\rho_{12}({\bf x},{\bf x}', t)=
{\rm e}^{- 2[C(0) + C(|{\bf x}-{\bf x}'|)]t/ \hbar^2 
}\rho^{(0)}_{12}.
\label{D72}
\end{eqnarray}
\noindent
The increase of the kinetic
energy, 
then, follows from Eq.(\ref{D5}) as

\begin{eqnarray}
\displaystyle
K=K(t)=-{N\over m}\nabla^2_{\bf x =0}C(|{\bf x}|)t,
 \quad K(t) = 
{ N C(0)\over m L_p^2}t
\label{D8}
\end{eqnarray}
\noindent
where the relations ~$\rho_{ii}({\bf x},{\bf x}, t)=
\rho^{(0)}_{ii},\,\, (\rho^{(0)}_{11}+ \rho^{(0)}_{22})V=N$,
with ~$V$~ being the total volume of the system,
have been employed.
We have also defined the noise 
correlation length ~$L_p$~ through the relation
 ~$\nabla^2_{{\bf x}=0}C(|{\bf x}|)
= - C(0)/L_p^2$, and imposed the condition
~${\bf \nabla}_{{\bf x}=0}C(|{\bf x}|)=0$.

We note that Eq.(\ref{D72}) accounts for the
loss of the correlations between the components
with the effective rate ~$\tau_{irr}^{-1}
\approx 2(C(0)+C({\bf x}-{\bf x}'))/\hbar^2$.
This rate can be obtained more accurately
from the equation
for the correlator
~$\rho_{12}({\bf x},{\bf x}', t,t'=0)$, which 
describes the decay of these correlations and carries
the information about the relative
phase (below $T_c$). Applying the Furutzu-Novikov theorem 
\cite{NOV}
again, and assuming that the initial state
is independent of the noise, we find

\begin{eqnarray}
\displaystyle
\rho_{12}({\bf x},{\bf x}', t, t'=0)=
{\rm e}^{- C(0)t/ \hbar^2 }\rho^{(0)}_{12},\,\, t>0.
\label{D9}
\end{eqnarray}
\noindent
Thus, the rate of the loss of the phase
memory is

\begin{eqnarray}
\displaystyle
\tau^{-1}_{irr}={C(0)\over \hbar^2},
\label{D10}
\end{eqnarray}
\noindent
which is a factor of 2-4 different
from the estimate obtained from the OPDM.

We define the time ~$\tau_h$~ as the time 
when the kinetic energy per 
particle becomes comparable to the
BEC transition temperature ~$T_c$ (so that
the BEC is destroyed).
Accordingly, employing Eq.(\ref{D8}),

\begin{eqnarray}
\displaystyle
\tau^{-1}_h\approx {C(0)\over
mT_cL^2_p}. 
\label{D11}
\end{eqnarray}
\noindent
Taking the ratio, we find

\begin{eqnarray}
\displaystyle
{\tau^{-1}_{irr}\over \tau^{-1}_h}\approx 
{mT_cL^2_p\over \hbar^2}=\left({L_p\over L_c}
\right)^2, 
\label{D12}
\end{eqnarray}
\noindent
where we have introduced the thermal
length ~$L_c=\hbar /\sqrt{mT_c}$~
at the transition temperature.

Thus, the loss of the phase memory
can occur faster than the destruction 
of the condensate, only if the 
correlation length ~$L_p$~ is longer
than ~$L_c$
\cite{DAMOP}. From general physical
considerations, it is clear that
this condition is not realized in
the actual traps, if the main source
of the noise is the background gas.
The correlation length ~$L_p$~ is simply
the thermal length of the hot gas, which
is obviously much shorter than
the thermal length of the confined 
cloud at ~$T_c$. 
It is also worth noting that similar
conclusion can be reached in the general 
case, when the random field produces
exchange between the components. Rigorous 
proof of this is also based on the Furutzu-Novikov
theorem. 

The losses from the trap
do not change the situation. The rate
of the decay of the phase memory 
is practically comparable to the
rate of the escape from the trap.
Physically, this is so because
the losses do not introduce any
long range correlations which
may disrupt the global relative phase
faster than the loss of particles. 
This conclusion
is based on the analysis of
the corresponding Wigner equations
following from the OPDM, where the
collision integral is derived from
the term (\ref{D13}). Detailed account of
this will be presented elsewhere.

It should be noted that a special attention
should be paid to the situation of 
the quasi-equilibrium, when, e.g., the 
deposition to the trap 
and the evaporative
cooling occur simultaneously, so that
the trap remains at constant temperature
and retains a fixed on average number of atoms.
In this case, both rates of the deposition
and of the losses can be large. 
Accordingly, the decoherence
will be fast despite that no significant
changes in the temperature and in the population
are observed. The analysis of this situation will
be presented elsewhere. 
Thus, {\it if all the individual
rates of disrupting the equilibrium
are kept low on the experimental time scale, the 
correlations between the BEC components 
will persist on this scale}.

Below we will analyze the situation
when the white noise is long ranged. As it will
be seen, the above conclusion does not hold 
in this case any more. 

\subsection{ The phase decoherence
induced by noisy external
long ranged potential}

The noisy potential ~$\xi$~ could be induced
by some external macroscopic fields. 
For example, some magnetic or electric
fields can be imposed by extended
macroscopic wires or by lasers.
These fields fluctuate due to thermal
and quantum effects. The magnetic trapping 
potential itself can fluctuate
due to the fluctuations of, e.g., 
electric currents in the magnetic
coils. The effect of the random 
term on the BEC dynamics has recently
been analyzed in, however,
different context in Ref.\cite{POT}.
In such situations,
the noisy field can be taken as
~$\xi ({\bf x}, t)=
U'({\bf x})\eta (t)$, where 
~$U'({\bf x}) $~ stands for some
macroscopic potential which breaks
the symmetry, and ~$\eta (t)$~
denotes the white noise factor
obeying the relation

\begin{eqnarray}
\langle \eta (t)\eta (t')\rangle
=\Delta \delta (t-t'),
\label{L1}
\end{eqnarray}
\noindent
where the spectral weight ~$\Delta$~
depends on the nature of the noise,
and will be evaluated later in the case
of the thermal equilibrium fluctuations.
The Heisenberg
equations for the two-component
BEC become (we again ignore
the self-interaction)

\begin{eqnarray}
\displaystyle
i\hbar\dot{\Psi}_{1,2}&=&
[-{\hbar^2 \nabla^2\over 2m} + U \pm  U'\eta]\Psi_{1,2},
\label{L21}
\end{eqnarray}
\noindent
where the "+"
sign is chosen for the first component, and
"-" -- for the second;  ~$U$~ stands for the
"symmetric" part of the trapping potential.

Employing Eqs.(\ref{L1}), (\ref{L21}), 
and applying the Furutzu-Novikov
theorem (see Appendix A ), we find
the equations for the OPDM

\begin{eqnarray}
\displaystyle
i\hbar\dot{\rho}_{ii}({\bf x},{\bf x}', t)&=&
\{{\hbar^2(\nabla^2_x -\nabla^2_{x'}) \over 2m}
-U({\bf x}) + U({\bf x}')
 - i{\Delta \over \hbar }
[U'({\bf x})-U'({\bf x}')]^2\}
\rho_{ii}({\bf x},{\bf x}', t);
\label{L31}
\\
\phantom{XXX}
\nonumber
\\
i\hbar\dot{\rho}_{12}({\bf x},{\bf x}', t)&=&
\{{\hbar^2(\nabla^2_x -\nabla^2_{x'}) \over 2m}
-U({\bf x}) + U({\bf x}')
 - i{\Delta \over \hbar }
[U'({\bf x})+ U'({\bf x}')]^2\}
\rho_{12}({\bf x},{\bf x}', t).
\label{L33}
\end{eqnarray}
We will employ the above equations
for estimating the 
magnitude of the ratio
of interest ~$\tau_{irr}/\tau_h$~
akin to Eqs.(\ref{D10})-(\ref{D12}).
The loss of the correlations between
the components is determined
by the term ~$\sim \Delta$~ in
Eq.(\ref{L33}), so that
the rate ~$\tau^{-1}_{irr}$~
can be estimated as 

\begin{eqnarray}
\displaystyle
\tau^{-1}_{irr}\approx 
{\Delta \over \hbar^2}\overline{U '^2}
\label{L4}
\end{eqnarray}
where the bar denotes the effective
averaging over the volume of the BEC. 

The heating rate can be found
from Eqs.(\ref{L31}). We calculate 
the rate of the increase of the total
energy in the trap as

\begin{eqnarray}
\displaystyle
&\dot{E}&= \sum_i\int \, d{\bf x}_{|x'=x}[
{-\hbar^2\nabla^2_{x'} \over 2m}
+ U({\bf x}')]
\dot{\rho}_{ii}({\bf x},{\bf x}', t)=
\nonumber
\\
\displaystyle
&\sum_i&\int \, d{\bf x}_{|x'=x}{\Delta \over 2m}
\nabla^2_{x'}(U'({\bf x})-U'({\bf x}'))^2\rho_{ii}
({\bf x}, {\bf x}',t)\approx N{\Delta \over m}
\overline{[{\bf \nabla}U']^2}
.
\label{L5}
\end{eqnarray}
\noindent
Accordingly, we find from Eq.(\ref{L5})
~$\tau^{-1}_h\approx {\Delta \over T_c m}
\overline{[{\bf \nabla}U']^2}$.
Then, the ratio ~$\tau^{-1}_{irr}/\tau_h^{-1}$~ becomes

\begin{eqnarray}
\displaystyle
{\tau^{-1}_{irr}\over \tau^{-1}_h}\approx 
{mT_c\overline{U '^2}
\over \hbar^2 \overline{[{\bf \nabla}U']^2}}.
\label{L50}
\end{eqnarray}
\noindent
If the potential is long ranged as, 
e.g., ~$U'\sim r^{\delta}$~ with
some $\delta >0$, and the cloud has
a typical size ~$r_c$, then,
~$\overline{U '^2}
/ \overline{[{\bf \nabla}U']^2}\approx r^2_c$.
Hence, we arrive at the estimate
Eq.(\ref{D12}), where the role
of the noise correlation length ~$L_p$~ is played
by the condensate typical size ~$r_c$.
Now it can be seen that, under the given
circumstances, ~$\tau^{-1}_{irr}/\tau_h^{-1}
\gg 1$, because ~$L_c \ll r_c$.
In other words, fluctuations of the
trapping potential (or any other extended
potential) destroy the inter-component
coherence much faster than producing any substantial
heating effect. This, however, does
not mean that the rate of the decoherence
is a significant issue in the actual traps,
if the noise of the trapping potential is
kept on the level of the thermal noise
even at room temperature. Indeed, let us estimate
the absolute value of the decoherence
rate (\ref{L4}). For this purpose, we
assume that the fluctuations ~$\eta U'$~ are
of the same order as those of the trapping potential
~$U$~ caused by the thermal noise ~$\delta I^{(e)}$~ of the
electric currents ~$I^{(e)}$~ which create the magnetic
trap. Then, we set ~$\eta 
\approx \delta I^{(e)}/I^{(e)}$, 
and
impose the condition of the white noise: 
~$\langle \delta I^{(e)} (t) \delta I^{(e)} (t')\rangle
=2T/R \delta (t-t')$,
following from
the classical limit of the fluctuation-dissipation
theorem.
Here ~$R$~ stands for the coil resistance.
Then we find ~$\Delta = 2T/RI^{(e)2} $~  
in Eq.(\ref{L1}). For the mean in Eq.(\ref{L4})
we employ 
~$\overline{U '^2}\approx \overline{U ^2}
\approx \mu_0^2$,
where ~$\mu_0$~ stands for the mean
chemical potential in the trap. Thus, Eq.(\ref{L4})
yields

\begin{eqnarray}
\displaystyle
\tau^{-1}_{irr}\approx 
{k_BT\mu_0^2 \over \hbar^2I^{(e)2}R},
\label{L40}
\end{eqnarray}
where we have reinstated the Boltzmann
constant ~$k_B$. Eq.(\ref{L40})
 gives ~$\tau^{-1}_{irr}=10^{-15}-10^{-13}$s$^{-1}$,
for typical
values ~$T= 300K,\,\, \mu/\hbar = 10^3-10^4$s$^{-1}$, and
for ~$I^2R=1$W. 
Thus, 
 the instrumental thermal noise
does not produce any significant decoherence.

\subsection{Intrinsic decoherence of the global relative
phase of the two-component BEC}

In this section we will discuss
mechanisms of the decoherence induced
by self-interaction in the trap. 
In principle, the two-component
condensate can be viewed as a system in which
certain information is imprinted into the relative
phase between the components. 
Any quantum phase memory device,
which could be employed in quantum 
information processing, should be able to retain
the phase information for long enough 
time.   
Thus, it is
important to understand the fundamental
mechanisms of the
relative phase memory loss.

A well studied mechanism of the
global phase decay at ~$T=0$~ - the phase diffusion
\cite{PD}
can be understood in terms of 
fluctuations of the chemical potential 
induced by binary interactions and 
by a finite variance of the number of particles
in the quantum {\it coherent} state.
These fluctuations produce collapse
of the global phase on some time
~$\tau_c$.
It has also been found that the PD 
exhibits spontaneous revivals \cite{PD}
on some revival time ~$\tau_R
\sim N \gg \tau_c$.
The time ~$\tau_R$, which is too
long on any experimental time scale,   can be considered
as a duration of the Poincare cycle. The spontaneous
revivals indicate that the dynamics is reversible
in a sense that it repeats itself. This, however,
does not necessarily mean that the dynamics of
~$N\gg 1$~ particles can be {\it physically}
reversed in time by imposing some physical 
external action.  
Thus,
it is important to understand the {\it physical
reversibility}
of the quantum dynamics of
the condensate on much shorter 
time scale   ~$\tau_c < t \ll \tau_R$. 
This question will be addressed later. 
We will show that the time ~$\tau_c$~
{\it is not} the time after which the phase memory 
is lost irreversibly. The phase 
information can be recovered (partially) by imposing 
the time reversal pulse.

Questions we are addressing here are:
Can the evolution of the global
relative be viewed as the PD process
at ~$T\neq 0$~? If it is the PD,
is this process physically reversible?  

Before we start, it is worth 
noting that 
 two conceptually very
different situations should
be clearly distinguished --  i) the
dynamics of the relative phase in the
presence of the permanent exchange of bosons 
between the components, and
ii) the phase dynamics with no exchange.
In the case i),
the relative global phase 
is characterized by some particular
equilibrium value. Deviations from it
increase the energy, and thereby activate
irreversible processes of 
relaxation to the equilibrium.
On the contrary, in the case ii),
if once
prepared due to a short exchange 
of bosons, the 
global relative phase {\it cannot} be 
viewed as a normal mode, if
no exchange of bosons exists 
during the following evolution (
akin to the situation realized by
the JILA group \cite{JILA}).We consider
the case ii) in this paper.

Here
we will extend our analysis Ref.\cite{OC}
of the problem of the global phase
time-correlations to the
two-component case. The central element
of this analysis is the concept 
of the {\it projected} Hamiltonian and
the {\it projected
many body eigenstates}, which will be often
called
just "projected states", of this Hamiltonian.
The total numbers of 
bosons in the each component play a role
of parameters (not the operators)
in such a Hamiltonian, if no losses
or exchange between the components
exist\footnote{The meaning of the 
projected Hamiltonian can be understood
from the analogy with
motion in a central potential, when
the total angular momentum (and its
$z$-component) is conserved.
Due to the separation of the angular variables,
the equation for the radial
part can be formulated in terms
of the Hamiltonian with some effective 
potential, which depends on the angular momentum
as a parameter. 
This Hamiltonian for the radial part 
is the {\it projected} Hamiltonian.}.

We represent an
exact many-body 
eigenstate ~$|m,N,M\rangle$, which
is characterized by the following
quantum numbers - the total number
of 
bosons ~$N=N_2 + N_1$~ and the half
of the population 
difference ~$M=(N_2 - N_1)/2$~ as well as 
by a set of the quantum numbers
~$m$~ referring to the normal
excitations, as an
expansion 

\begin{eqnarray} 
\displaystyle |m,N,M\rangle &=& \sum_{n_{\alpha 1},n_{\alpha 2},...
}
C_{n_{\alpha 1},n_{\alpha 2},...
}(m,N,M)
|n_{10}\rangle |n_{20}\rangle |n_{11}\rangle 
|n_{21}\rangle ... \,\, ,
\label{16}
\\ 
n_{\alpha 0}&=&N_{\alpha}
-(n_{\alpha 1}+n_{\alpha 2}+...), 
\nonumber  
\end{eqnarray}
\noindent
in the Fock space ~$|n_{10}\rangle 
|n_{20}\rangle |n_{11}\rangle 
|n_{21}\rangle ...  $~ 
of the population numbers
~$n_{\alpha j}$~ of some set of single particle
states. Here the Greek index refers to the
component ~$\alpha = 1,2$, and the Latin
index labels the single particle states
in the corresponding component, so that
~$n_{\alpha 0}$~ stands for the
population of the ~$\alpha$th condensate 
component.

The expansion coefficients
~$ 
C_{n_{\alpha 1},n_{\alpha 2},...}(m,N,M)$~ 
form the Fock 
representation of the eigenstates.
We call these coefficients
the {\it projected states}, and introduce
a short notation ~$|\widehat{m,N,M\rangle }$~ 
for them \cite{OC}. 

It is important to note that, while being
orthogonal with respect to the set ~$m$~
for given ~$N,\, M$, the projected
states {\it are not} orthogonal
for different ~$N,\, M$. In other words,
~$\widehat{\langle M, N, m'}|
\widehat{m,N,M\rangle }=\delta_{m'm}$~
and 
~$\widehat{\langle M', N', m}|
\widehat{m,N,M\rangle }\neq \delta_{M'M}
\delta_{N'N}$. Furthermore, in the case
of the exact ~$su(2)$~ symmetry,
the exact relation ~$
\widehat{\langle M', N, m}|
\widehat{m',N,M\rangle }=\delta_{m'm}$~
holds regardless of the values ~$M,M'$.

In 3D systems, close to equilibrium,
the population factor of the condensate state
in the each component is macroscopic, and
is characterized by relatively small
fluctuations. If
~$\Delta n_{\alpha 0}$~ stands for the
fluctuation of the condensate
population, and ~$\overline{n}_{
\alpha 0}$~ is the mean population,
the relative value ~$\Delta n_{\alpha 0}/
\overline{n}_{\alpha 0} \ll 1$~ \cite{FLUK}.
This circumstance can be employed 
to express matrix elements of any condensate operator
in terms of the overlap of the 
projected states. Specifically,
let us say one wishes to find the
matrix element ~$(a^{\dagger}_{20}
a_{10})_{m,m'}=
\langle M, N, m|a^{\dagger}_{20}
a_{10}|m',N,M-1\rangle$~ of the
condensate operators. Employing
the representation (\ref{16})
and the smallness of the fluctuations
of the condensate state populations,
one finds

\begin{eqnarray}
\displaystyle (a^{\dagger}_{20}
a_{10})_{m,m'}&=&
\sum_{n_{\alpha 1},n_{\alpha 2},..}
\sqrt{n_{10}(n_{20}+1)}C^*_{n_{\alpha 1},n_{\alpha 2},...}
(m,N,M)
C_{n_{\alpha 1},n_{\alpha 2},...}
(m',N,M-1) 
\nonumber
\\
\phantom{XXXX}
\nonumber
\\
&=&\sqrt{\overline{n}_{10}\overline{n}_{20}}
[\widehat{\langle M, N, m}|
\widehat{m',N,M-1\rangle } + o\left({\Delta n_{\alpha 0}
\over \overline{n}_{\alpha 0}}\right)].
\label{30}
\end{eqnarray}
\noindent
Thus, the matrix elements of the condensate
operators are given by the overlap of the
corresponding {\it projected}
eigenstates, with different values
of the population difference.
This simplifies
considerably calculations
of the correlators of the
condensate operators. 

Indeed,
the global relative phase
time-correlation properties are
completely described by
the correlator 
~$\rho_{12}(t)=\langle
a^{\dagger}_{10}(t)a_{20}(t) a^{\dagger}_{20}(0)a_{10}(0)
\rangle$. Employing Eq.(\ref{30}),
one obtains

\begin{eqnarray}
\displaystyle
\rho_{12}(t)=\sqrt{\overline{n}_{10}\overline{n}_{20}}
\langle {\rm e}^{iH(N,M)t} {\rm e}^{-iH(N,M-1)t}
\rangle
\label{I1}
\end{eqnarray}
\noindent
within the chosen accuracy (here and below
we employ units in which ~$\hbar =1$). 
$H(N,M)$~ stands for the {\it projected}
Hamiltonian defined as ~$H(N,M)=\sum_m \widehat{|m,N,M\rangle}
E_m(N,M)\widehat{\langle M,N,m|}$~, with
~$E_m(N,M)$~ being the exact many body eigenenergies
\cite{OC}.
The averaging is performed
over the thermal ensemble. 
Explicitly, 
~$\langle ... \rangle =\sum_{m,N,M}P_m(N,M)
\widehat{\langle M,N,m|} ...
\widehat{|m,N,M \rangle}$, where
~$P_m(N,M)$~ stands for the canonical
Boltzman factor.

The evolution operator
~$ {\rm e}^{-iH(N,M-1)t}$~ can be expressed
 as ~${\rm e}^{-iH(N,M-1)t}=
{\rm e}^{-iH(N,M)t}{\rm Texp} (-i\int^t_0\, dt' H'(t'))$~ 
, where the zeroth-order part
is ~$H(N,M)$~ and the perturbation
~$H'=H(N,M-1) - H(N,M)$~ is taken in the
interaction representation with respect to
the zeroth-order part ~$H'(t)= {\rm e}^{iH(N,M)t}H'
{\rm e}^{-iH(N,M)t}$. Finally,
Eq.(\ref{I1}) takes a form

\begin{eqnarray}
\displaystyle
\rho_{12}(t)&=&\sqrt{\overline{n}_{10}\overline{n}_{20}}
\langle {\rm Texp}(-i\int^t_0\, dt' H'(t'))
\rangle, 
\label{I20}
\\
 H'(t)&=& 
{\rm e}^{iH(N,M)t} \{
-{\partial H(N,M) \over \partial M} +
{1\over 2} {\partial^2 H(N,M) \over \partial M^2} + ...
\} {\rm e}^{-iH(N,M)t}
\label{I2}
\end{eqnarray}
\noindent 
within the chosen accuracy. 
It is clear that
any correlator of the condensate operators
can be represented in the form similar
to Eq.(\ref{I20}), (\ref{I2}).  

In fact, such an expansion is valid as long as
a macroscopically populated
BEC is present. Otherwise, the projected
Hamiltonian is a very sharp function of ~$M$, because
a transfer of even a single atom between the thermal
components inevitably corresponds to creating 
excitations in the case of the broken
~$su(2)$~ symmetry. 
In the presence of the BEC,
the 
each derivative term in the expansion of ~$H'(t)$~
introduces an additional factor ~$1/N$.
Accordingly, it is enough to keep
the first term only, so that ~$H'={\partial H(N,M)\over
\partial M}$. ~$H'$~ can be viewed as
an {\it operator} of the relative chemical
potential
~$\mu_m=
E_m(N,M) - E_m(N,M-1)\approx
\partial E_m(N,M)/\partial M$~ in the given
eigenstates.
The diagonal matrix
elements of ~$H'$~ give the {\it exact} values of the
~$\mu_m$.
The ensemble mean of it, then,
becomes ~$\overline{\mu} = \sum_{m,N,M}P(m,N,M)\mu_m=
\langle \partial H(N,M) 
/\partial M\rangle $.
It is important that
~$H'=0$, if the intrinsic symmetry holds. Thus, another
interpretation, ~$H'$~ is proportional to the change of the
interaction energy per one transferred particle.
As discussed in Ref.\cite{OC}, the proportionality
factor is of the order of unity in the one-component
BEC. In the two-component situation, this factor
should contain the combination
of the scattering lengths which vanishes at the
point of the intrinsic ~$su(2)$~ symmetry. 

We employ   
the cumulant expansion
of ~$\widehat{\langle M,N,m|} {\rm Texp}(
-i\int^t_0\, dt' H'(t'))
\widehat{|m,N,M \rangle}$. Then, within the accuracy
~$1/N$, we reproduce Beliaev's result \cite{BEL}
~$\rho_{12}(t)\sim \exp (i \overline{\mu}t)$.
We are, however, interested in the next non-vanishing
effect of large but finite ~$N$. Therefore, the next 
cumulant should be retained as well, and the higher
ones can safely be neglected 
within the ~$1/N^2$~ accuracy. 
Accordingly, we write

\begin{eqnarray}
\displaystyle
\widehat{\langle  M,N,m|}{\rm Texp}(-
i\int^t_0\, dt' H'(t'))\widehat{|m,N,M \rangle}&=&
\nonumber
\\
\nonumber
\\
\exp \{i\mu_mt- \int^t_0\, dt'\int^{t'}_0\, d\tau
\widehat{\langle M,N,m|}&\delta& H'(\tau) \delta H'(0)
 \widehat{|m,N,M \rangle} + o(1/N^2)\},
\label{I3}
\end{eqnarray}
\noindent
where we have introduced the notation
~$ \delta H'(\tau)$~ for the off-diagonal
part of the total matrix ~$H'(\tau)$. The
operator ~$ \delta H'(\tau)$~ can be viewed
as fluctuation of ~$H'$.

As it has been discussed in Ref.\cite{OC},
the double integral in the exponent determines
the processes when adding or removing one boson
to or from the BEC disturbs the normal component.
In principle, a situation can be contemplated
when such a disturbance is so strong
that this term dominates, and rapidly
suppresses the overlap of the projected
states (differing in the total number of bosons
by 1). This situation is similar to
the orthogonality catastrophe (OC)
occurring in Fermi liquids \cite{EDGE}.
In Beliaev's work \cite{BEL}, however,
 it has been proven
that such processes are insignificant at ~$T=0$.
Thus, no the OC is to be anticipated
in the BEC at ~$T=0$. This implies that
the double integral in Eq.(\ref{I3})
is finite and small in the limit ~$t\to \infty$.
 
At finite ~$T$,
without a rigorous proof, the Beliaev's result
is widely applied to higher eigenstates \cite{STAT}.
Below, we will employ Eq.(\ref{I3}) and give
semi-quantative general arguments which support
the Beliaev's result at finite ~$T$.  
First, we note that the correlator
in Eq.(\ref{I3}) can be calculated within
the perturbation theory with respect to
interaction between the quasiparticles.
This has been done in the case of the
single-component BEC in Ref.\cite{OC}.
The result of such calculations  can
be represented as 

\begin{eqnarray}
\displaystyle
\sum_{m,N,M}P(m,N,M)\widehat{\langle M,N,m|}{\rm Texp}\{-
i\int^t_0\, dt' H'(t')\}\widehat{|m,N,M \rangle}=
{\rm e}^{i\overline{\mu}t - (t/\tau_d)^2 -t/t_{OC}},
\label{I4}
\end{eqnarray}
\noindent
where

\begin{eqnarray}
\displaystyle \tau_d^{-2}={1\over 2} \langle (\mu -
\overline{\mu})^2\rangle
, \quad
t^{-1}_{OC}\approx \int^{\infty}_0 \, dt
\langle \delta H'(t) \delta H'(0)\rangle.
\label{I5}
\end{eqnarray}
\noindent
The gaussian factor in Eq.(\ref{I4}) is a result
of applying the central limit theorem for the averaging
of ~${\rm e}^{i\mu_mt}$~ over the ensemble.
The quantity ~$t_{OC}$~ has been called
the OC time \cite{OC}. 
It determines how rapidly the overlap 
decays. Its decay is controlled by 
excitations. The rate ~$t^{-1}_{OC}$~       
can be estimated 
as 

\begin{eqnarray}
\displaystyle
t_{OC}^{-1}\approx 
\langle \delta H'^2\rangle \tau_n
\label{I50}
\end{eqnarray}
\noindent
where ~$\tau_n$~ stands
for a typical relaxation time
of the correlator  
~$\langle \delta H'(\tau) \delta H'(0) \rangle$
which is taken in the exponential form
~$\sim \exp (-t/\tau_n)$.
In 3D system the time ~$\tau_n$~ is finite because of very
small statistical weight of the low energy modes.
This time is given by a typical relaxation 
time of the normal modes characterized by
the energies around the chemical
potential. 
Obviously, ~$\tau_n \ll \tau_d$~ (\ref{I5}).
Thus, taking into account the nature of the
operator ~$H'$~ discussed above (see below Eq.(\ref{I2})),
we conclude that ~$t_{OC}^{-1}\sim 1/N$, because
~$\langle \delta H'(0) \delta H'(0) \rangle \sim 1/N$
as fluctuation of 
any extensive quantity taken per one particle.
It should, however, be noted that the form
(\ref{I4}) {\it is not } valid in the limit
~$t \to \infty$. 
In Ref.\cite{OC}, we have discussed that
the double integral in Eq.(\ref{I3})
should remain constant and small in the limit
~$t=\infty$ (~$t\geq t_{OC}$) even at
~$T\neq 0$. 
The reason for this is the following:
the spectral weight of the correlator
~$\langle \delta H'(t) \delta H'(0)\rangle$~
in the limit of zero frequencies
is effectively collected from the lowest order
processes
of scattering of the quasiparticles
with almost zero energy transfer. 
The term ~$H'$, which
is actually proportional to the interaction
energy, removes the degeneracy between the
corresponding quasiparticle states. This
 suppresses the spectral weight
in a very similar manner to the effect
of the level repulsion studied in the
Random Matrix Theory \cite{RMT}. 
This is essentially a non-perturbative effect, and
it becomes important at frequencies
smaller than a typical value of the interaction
matrix element, which is scaled as ~$\sim 1/\sqrt{N}$~
\cite{OC}. Thus, the exponential factor
~${\rm e}^{-t/t_{OC}}$~ in Eq.(\ref{I4}) is 
correct only for times shorter than some time
~$t'\sim \sqrt{N}$, and it levels off to a constant
~$\sim \exp (-t'/t_{OC}) \approx 1 + o(1/\sqrt{N})$~
(note that ~$t_{OC} \sim N$~) as ~$t$~ becomes longer
than ~$t'$.

The quantity ~$\tau_d$~  is determined
by the ensemble fluctuations
of the chemical potential, and it describes
the rate of the phase diffusion at
finite temperature. We will call it the
thermal PD in order to distinguish it from
the quantum PD \cite{PD}.
Thus,
~$\tau^{-1}_d \sim 1/\sqrt{N} \gg t^{-1}_{OC} \sim 1/N$,
and, therefore, the OC is irrelevant
because 
the reversible dephasing dominates in Eq.(\ref{I4})
on any practical time scale, provided no 
extrinsic factors play any role.

The evolution
of the condensate operators can be viewed
as a result of the ensemble averaging
of the {\it non-decaying}
exponents, each representing a particular
realization of the chemical potential.
This is so, if the relation

\begin{eqnarray}
\displaystyle
\widehat{\langle M,N,m }
\widehat{|m',N,M \pm 1\rangle} =
\delta_{m,m'}
\label{I6}
\end{eqnarray}
\noindent
holds.
We note that Eq.(\ref{I6})
is, in fact, an extension \cite{STAT}
of the Beliaev's result \cite{BEL}\footnote{ that
the action of the condensate operators on the
ground state does not produce excitations 
within ~$1/N$~ accuracy} for finite $T$
formulated in terms of the overlap 
of the projected states. 
Accordingly, the correlator (\ref{I1})
simply becomes 

\begin{eqnarray}
\displaystyle
\rho_{12}(t)=\sqrt{\overline{n}_{10}\overline{n}_{20}}
\sum_{m,N,M} P(m,N,M){\rm e}^{i\mu_m t}=
\sqrt{\overline{n}_{10}\overline{n}_{20}}
\,{\rm e}^{i\overline{\mu}t - (t/\tau_d)^2}
\label{I7}
\end{eqnarray}
\noindent
within the central limit approximation.

The dephasing time ~$\tau_d$~ can be calculated
in any order with respect to the interparticle
interaction. For this purpose, in the 
weakly interacting system,
it is enough to consider the Bogolubov 
gas of the non-interacting quasi-particles
as a model of the excited states. Specifically,
the eigenenergies are represented as ~$E_m(N,M)=
E_0(N,M) + \sum_a \epsilon_a(N,M) n_a$, where
~$E_0(N,M)$~ is the ground state energy and 
~$\epsilon_a(N,M)$~ stands for the spectrum of
the quasi-particles characterized by the population
factors ~$n_a$. Then, Eq.(\ref{I5}) yields

\begin{eqnarray}
\displaystyle \tau_d^{-2}&=&
{1\over 2}\langle \left(\sum_a
{\partial \epsilon_a(N,M)
\over \partial M}n_a - \langle \sum_a
{\partial \epsilon_a(N,M)
\over \partial M}n_a\rangle \right)^2\rangle
+ \tau_c^{-2},
\label{I8}
\\
\tau_c^{-2}&=& {1\over 2}
\langle\left(
{\partial E_0(N,M)\over \partial M}
-\langle {\partial E_0(N,M)\over 
\partial M}\rangle \right)^2\rangle,
\label{I9}
\end{eqnarray}
\noindent
where ~$\tau_c$~ stands for the quantum
phase diffusion (at ~$T=0$) collapse time.
The quantum PD has been analyzed by many
authors in Refs.\cite{PD} in the case
when the initial state is the {\it coherent}
state with well defined global phase.
We note that, in the 
two-component BEC, the shot noise
in the initial deposition of 
~$N$~ should also contribute
to the dephasing of the
condensate correlators at ~$T=0$ even in the
situation when no initial global relative
phase existed.
Furthermore, in actuality, the shot noise 
is most likely
to wash out completely the effect of the quantum
PD in Eq.(\ref{I9}). 
In our work \cite{SU2}, we discussed
this in detail, and have found 
that in ~$^{87}$Rb the collapse time
 ~$\tau_c$~ can be as short as 30ms,
if the shot noise reaches values about
10\%. This should be compared with
typical estimates of the quantum PD
collapse times ~$\sim 1-10$s \cite{PD}.

It should also be noted
that ~$\tau_d^{-1}=0$~ in the case
of the exact $su(2)$ symmetry.
Here we will not focus on calculating
the dephasing rates for the
specific geometries \cite{JILA}, and will use it as 
a parameter.

Summarizing, the evolution of the condensate
operators does not disturb the normal component,
and, as a consequence, 
the correlators of the condensate operators
exhibit reversible dephasing determined by
the ensemble fluctuations of the relative
chemical potential. In the next section we will
discuss how such a reversibility can be tested 
in the atomic echo experiment. 

\section{Atomic echo}

As discussed above, the evolution of the 
condensate operators in 3D system
including finite number of bosons
does not disturb significantly
the normal component. 
The decoherence of the condensate operators
is a result of the ensemble averaging, with
no decoherence occurring in any eigenstate.
In this section,
we will address the question
of 
{\it physical
reversibility}, so that an initial
phase information can be recovered
on times longer 
than the dephasing
time ~$\tau_d$. 

In general, the echo effect (spin echo,
photon echo \cite{ECHO}, plasma echo \cite{PLASMA}) --
stimulated revival of some quantity --
 is widely employed as a test
for reversibility of dynamics.
Once initially prepared, the quantity
may exhibit an apparent decay, which is 
a sort of inhomogeneous broadening
without actual loss of the memory of the
initial state. Then, if some time-reversal
pulse is imposed at later time
~$\cal{T}$~, which can be
much longer than the dephasing time ~$\tau_d$~,
the quantity may revive completely or
partially at time ~$t\approx 2\cal{T}$, and,
if the time reversal pulse is strong,
at integers of ~$\cal{T}$ (multiple echoes).

Generally speaking, any Hamiltonian dynamics
is reversible in time. This means
that the formal time-reversal of the evolution 
will result in the restoration of the
initial state. This, however, cannot 
always be done physically, that is,
by applying some physical disturbance to
the system. Thus, a proper criterion
for the {\it physical reversibility}
of the dynamics should be introduced. 
We will employ the criterion  which relies on the
strength of the echo. Specifically:

{\it If an external pulse can induce
the echo whose strength
is significantly 
larger than fluctuations (quantum or statistical)
of the measured quantity, the evolution of this
quantity can be considered (partially) reversible}.

It is important to realize that the echo in 
the condensate
can be observed even though the time delay ~$\cal{T}$~
is much longer than typical relaxation times of
the excitations in the system and/or in the
bath. The time delay
should be shorter than the irreversibility time
~$\tau_{irr}$~ (\ref{D10}), (\ref{L4}).
As discussed in Sec. III, this time is 
determined by the strength of the extrinsic
fluctuations, which can even be treated
within the $\delta$-correlated approximation,
that is, within the assumption of {\it infinitely
short} relaxation times.

In our work \cite{SU2}, we have presented the echo
solution in the two-component BEC in the case
of the shot noise dominated dephasing.
In this situation, the echo
strength reaches 100\%, and is achieved by imposing
the ~$\pi$-type time reversal pulse. Here we  
will analyze the echo in the case of the
intrinsic dephasing, that is, when either
the quantum or thermal PD takes place.

\subsection{Atomic echo at ~$T=0$}

Let us consider the echo 
experiment in the context of the
three pulses scheme \cite{SU2}.
We take ~$U_1$~ as the evolution operator
which corresponds to the ~$\pi /2$~ pulse
which converts the initial state with
 ~$N_2=N,\,\,\, 
N_1=0$~ into
a coherent state of the two condensates
\cite{JILA}. Then, some time-reversal pulse ~$U'$~ 
is imposed at time ~${\cal T} >0$~ later,
and, finally, the read out ~$\pi/2$~
pulse \cite{JILA} is 
imposed at time ~$t>\cal{T}$~. 

We choose, as a measured quantity, 
the population difference represented
by the operator ~$I_z$~ in eq.(\ref{3})
and measured by the JILA group
\cite{JILA}. A state vector
is now labeled as ~$|j, M\rangle$~
by the conserving total angular
momentum ~$j=N/2$~ and by its $z$-projection
$M$. At ~$T=0$, the excitation label
"$m$" can be suppressed.
Then,
the mean after the read out pulse becomes

\begin{eqnarray}
\displaystyle \langle I_z\rangle= 
\langle j,j|U_1{\rm e}^{i{\cal T} H}
(U')^{\dagger}
{\rm e}^{i(t-{\cal T})H}U_1^{\dagger}
J_zU_1{\rm e}^{-i(t-{\cal T})H}U'
{\rm e}^{-i{\cal T} H}U^{\dagger}_1
|j,j\rangle\,\, .
\label{90}
\end{eqnarray}
\noindent
The pulse operators ~$U_1,\,\, U'$~ can be 
found in the sudden approximation, if
the duration of the pulses is short.
Then, a general pulse operator is
~$U(|v|, \phi)=\exp(-i(v^*I_++ H.c.))$~, where
~$v_+=\int dt \Omega(t)=|v|\exp(i\phi)$~ 
is the time integral (over the pulse duration)
of the
Rabi frequency ~$\Omega (t)$~ in Eq.(\ref{0}),
and the operators ~$I_{\pm}$~ are represented
in Eq.(\ref{3}). The value ~$|v|$~ is the 
magnitude of the pulse, and ~$\phi$~ stands 
for its phase. Specifically, for the
~$\pi /2$-pulse, one can choose ~$U_1=U(\pi/4,
\phi=-\pi/2)=\exp(i{\pi \over 2}I_y)$. For the
time reversal pulse, we will take
~$U'=\exp (i\beta I_y)$, where ~$\beta$~
stands for the pulse strength. The dependence
of the echo strength on ~$\beta$~ will be
investigated below.

It is worth noting that the operator 
~${\rm e}^{i \beta I_y}$~ is the operator
of finite rotations, and ~$\beta$~
stands for the corresponding Euiler angle. 
An explicit form for the matrix elements
~$d^{(j)}_{M'M}(\beta)=\langle M', j|
{\rm e}^{i\beta I_y}|j, M\rangle$~
is well known for any
arbitrary dimension of the ~$su(2)$~ representation
(see, e.g., in Ref.\cite{LAN}). 
For example, 

\begin{eqnarray}
\displaystyle d^{(j)}_{jM}(\theta)= 
\langle j,j|{\rm e}^{i\theta
I_y}|j,M\rangle 
=\sqrt{ {(2j)!\over (j+M)!(j-M)!}}
\left(\cos {\theta \over 2}\right)^{j+M}
\left(\sin {\theta \over 2}\right)^{j-M}
\label{J1}
\end{eqnarray}
\noindent
where ~$M=-j, -j+1, ...,j-1, j$. The
expression for arbitrary ~$M,M'$~ involves
Jacobi polynomials, and can be found 
in Ref.\cite{LAN}.

Thus,
the mean (\ref{90}) can be calculated
as a finite product of the matrices ~$N+1$ by
~$N+1$.
In our work \cite{SU2}
we have analyzed the exact echo solution,
when the main source of the dephasing
is the shot noise.
In what follows, we study the echo
effect in the situation of the
quantum PD \cite{PD}. As it can be seen
later, the echo solution can be found,
practically, exactly 
as well. 

First, we remind that, under the given
initial condition, the
evolution of the population difference
represented by Eq.(\ref{90}) exhibits
collapses and spontaneous revivals
\cite{PD}. Indeed, setting the time reversal
pulse to zero in Eq.(\ref{90}), one finds

\begin{eqnarray}
\displaystyle 
\langle I_z \rangle=
\langle j,j|{\rm e}^{i{\pi \over 2} J_y}{\rm e}^{itH}
J_x{\rm e}^{-itH}{\rm e}^{- i{\pi \over 2}J_y}
|j,j\rangle,
\label{J3}
\end{eqnarray}
\noindent
where we have employed
the standard $su(2)$ relation 
~${\rm e}^{- i{\pi \over 2} I_y}I_z{\rm e}^{ i{\pi \over 2}
 I_y} =I_x$. 
Taking into account Eqs.(\ref{J1}), and the
dependence of the chemical potential on the
population difference,
one immediately obtains the collapses revivals
solution \cite{PD}, where the collapse
rate is ~$\tau_c^{-1}$~ (\ref{I9}) 
is a factor of ~$\sqrt{N}$~ larger than
the revival rate ~$\tau_R^{-1}$~. 
In what follows, we will assume that
~$N\gg 1$, so that the revival time is too long
on any practical time scale. 
Furthermore, as we mentioned above,
even weak shot noise
should wash out completely any spontaneous revivals.
Thus, we will be always assuming that 
~$\tau_c < \cal{T} \ll \tau_R$,
so that the echo - {\it stimulated} revivals would
not be confused with the {\it spontaneous}
revivals.

Calculations of the mean (\ref{90}) exactly
for large $j$ and for ~$\beta \neq 0$~ is a very
combersome problem. We simplify it
by noting that, as indicated
by Eq.(\ref{J1}), the initial ~$\theta =\pi/2$-pulse 
creates states with ~$|M| \approx \sqrt{j}
\ll j$. Accordingly, the time reversal
pulse ~$U'$~ acts effectively between states  
with such ~$M$. This, implies that the
formalism of the quantum relative phase
can be applied for calculating the matrix
elements of ~${\rm e}^{i\beta I_y}$.
We introduce the phase ~$\varphi$~
as a conjugate variable to ~$I_z$. 
Thus, the representation
of the operators (\ref{3}) 
can be chosen as

\begin{eqnarray}
I_z=- i {\partial \over \partial \varphi }\,\, , \quad
I_x=j \cos \varphi + {1\over 2j} 
{\partial \over \partial \varphi } \cos \varphi
{\partial \over \partial \varphi } + o(1/j^3),\,\,\ 
\nonumber
\\
\phantom{XXXX}
\label{P3}
\\
I_y=j \sin \varphi + {1\over 2j} 
{\partial \over \partial \varphi } \sin \varphi
{\partial \over \partial \varphi } + o(1/j^3),\,\,\ 
\nonumber
\end{eqnarray}
\noindent
which is a formal expansion in ~$1/j=2/N$. The terms 
shown explicitly satisfy
the standard ~$su(2)$~ commutation relations within
the accuracy ~$1/N^2$. We emphasize that
the quantum phase representation is valid only on the
space of the
states characterized by ~$|M|< \sqrt{j}$. 
As it will be seen later, the maximum echo can be achieved
for the strengths of the pulses which satisfy
the relation ~$\beta \sim 1/\sqrt{ j}$. Accordingly,
within the formal series (\ref{P3}), ~$\beta I_y
\approx \beta j\sin \varphi$~ and ~${\rm e}^{i\beta I_y}
\approx {\rm e}^{i\beta j\sin \varphi}$, so that
the matrix elements 
become

\begin{eqnarray}
d^{(j)}_{M'M}(\beta) \approx J_{M-M'}(\beta j),
\quad |M|, |M'| \ll \sqrt{{j\over \beta}}, \,\, j\gg 1,
\label{N1}
\end{eqnarray}
\noindent
where
~$J_n(x)$~ is the Bessel function of the first
kind of the ~$n$th order.
This relation can also be obtained
from the hypergeometric series, which
yields both the
limiting representation of the Bessel functions,
and the Jacobi polynomials \cite{ABRAM}.
 This approximation simplifies
calculations of the mean (\ref{90}) considerably.
On Fig.1, we have compared the 
results of the exact numerical
calculations of (\ref{90}) with the calculations
within the quantum phase method. As can be seen, agreement
is very good (for ~$N\approx 100$), and it 
is naturally expected to become better as ~$N$~ increases.

Accordingly, in what follows we will be
discussing the quantum phase method only. 

Thus, the complete representation
of the mean (\ref{90}) becomes

\begin{eqnarray}
\displaystyle 
\langle I_z \rangle &=&2{\rm Re}
\sum_{M,M',K} d^{(j)}_{jM}({\pi \over 2}) d^{(j)}_{M'j}(
-{\pi \over 2})(I_x)_{KK-1}
J_{K-M}(\beta j)
J_{K-M'-1}(\beta j)\times
\nonumber 
\\
&{\rm e}&^{ i[(E_0(M)-E_0(M')){\cal T} +
(E_0(K)-E_0(K-1))(t-{\cal T})]},
\label{J4}
\\
E_0(M)&=&(\epsilon_0 +b_1N)M +{b_2\over 2}M^2
+{b_3\over 2}N^2,
\label{J5}
\end{eqnarray}
\noindent
where ~$(I_x)_{KK-1}=(I_x)_{K-1K}={1\over 2}
 \sqrt{(j+K)(j-K+1)}$~ stands for the matrix 
elements of the operator ~$I_x$;
we have 
introduced the ground state
eigenvalue ~$ E_0(M)$~
of the Hamiltonian (\ref{0}) as a function
of the half of the population
difference ~$M$~ and of the total
number of bosons ~$N$, with ~$b_{1,2,3}$~ being
the coefficients which can be obtained from
Eq.(\ref{0}) within, e.g., the two-mode
approximation \cite{TWO}. 
The results of the numerical
calculations of Eq.(\ref{J4}) are presented on Fig.2
(for ~$b_1=0$).
As can be seen, the multiple echoes occur at
times ~$t\approx k{\cal T}, \,\, k=2,3,4,...$.

The multiple echo solution can be found, 
practically, exactly
in the limit ~${\cal T} \gg \tau_c =2
 (\sqrt{j}b_2)^{-1}$. 
As it will be seen later, the maximum
echo is achieved when ~$\beta \approx \tau_c /({\cal T}
\sqrt{j})$, so that the argument of the Bessel functions
is ~$\approx \sqrt{j}\tau_c/{\cal T} \ll \sqrt{j}$.
The Bessel function ~$J_{n-m}(z)$~ becomes essentially zero
when its order ~$|n-m| > |z|$~ \cite{ABRAM}. Thus,
in Eq.(\ref{J4}), while the index ~$M$~ runs over
the range ~$\approx [-\sqrt{j}; \,\, \sqrt{j}]$,
the other indices stay relatively close to ~$M$.
Accordingly, in the sum  (\ref{J4}),
we approximate ~$\sqrt{(j+K)(j-K+1)}\approx j$.
Then, the sum over ~$K$~ can formally be extended 
from ~$-\infty$~ to ~$+\infty$. This, allows
to calculate this sum in a closed form
by employing Eq.(\ref{J5}) and by implementing
the identity (9.1.79) of Ref.\cite{ABRAM}.
Accordingly, Eq.(\ref{J4}) acquires the form

\begin{eqnarray}
\displaystyle 
\langle I_z \rangle =j\sum_{L=1}
J_{L+1}(\beta jb_2(t-{\cal T}))
{\rm e}^{-(t - {\cal T}(L+1))^2/\tau_c^2 }
\cos [\epsilon_0(t-{\cal T}(L+1)) + {\pi (L+1)\over 2}],
\label{J6}
\end{eqnarray}
\noindent
where we have introduced the notation ~$L=M'-M$, 
and have eliminated all the terms with
~$L\leq 0$~ because their contributions
become significant only at negative times.
We also employed
the central limit theorem and made
the replacement
~$\sum_M d^{(j)}_{jM}({\pi \over 2}) 
d^{(j)}_{M+Lj}(-{\pi \over 2})
\exp [ ib_2(M+L/2)(t-{\cal T}(L+1))]
\approx \exp [- (t - {\cal T}(L+1))^2/\tau^2_c]$,
which is valid for ~$L \ll \sqrt{j}$. 
  
Eq.(\ref{J6}) describes the stimulated revivals --
the echoes. They occur at the times ~$t\approx
{\cal T}(L+1), \,\, L\geq 1$. The analytical
solution (\ref{J6}) is shown on Fig.2 together
with the numerical one, both calculated within
the quantum phase method. The agreement is quite
good.

It is interesting
to note that the magnitude of the
echoes exhibits non-monotonous  dependence
on the strength ~$\beta$~ of the time-reversal pulse.
Another feature is that the maximum
echo can be created by the {\it weak }  
time-reversal pulse. Let us consider this in detail
for the first echo ($L=1$) at the maximum
of the gaussian, which occurs at ~$t=2\cal{T}$. Then,  
the relative  echo magnitude becomes

\begin{eqnarray}
\displaystyle 
{|I_z| \over j}= 
J_2(\beta j b_2 {\cal T})= 
J_2(2\beta \sqrt{j}{\cal T}/\tau_c),
\label{J7}
\end{eqnarray}
\noindent
where we have expressed ~$b_2$~ in terms 
of the collapse time, which follows
from Eqs.(\ref{I9}), (\ref{J5})
as ~$\tau_c^{-1}=\sqrt{j}|b_2|/2$~
(we ignore the shot noise). Taking into account
that the first maximum ~$J_2(z)\approx 0.5$~ occurs at
~$z\approx 3$~ \cite{ABRAM}, we find the magnitude of the
strength of the time-reversal pulse

\begin{eqnarray}
\displaystyle 
\beta_{max}\approx 1.5 {\tau_c \over {\cal T} \sqrt{j}}
\label{J8}
\end{eqnarray}
\noindent
required to achieve the maximum echo.
Conversely, ~$J_2(z)=0$~ for ~$z\approx 5$
for the first time, and
the echo vanishes for ~$\beta \approx 1.7\beta_{max}$.
Taking the higher value of ~$\beta$~ results
in the oscillatory dependence of the echo
strength on ~$\beta$, with the total amplitude
slowly diminishing as ~$\sim 1/\sqrt{\beta}$. 
Specifically, for ~$\beta \sim 1$~, Eq.(\ref{J7})
yields ~$|I_z|/j \sim j^{-1/4} \ll 1$. As it will be seen
later, at ~$T\neq 0$~ this dependence can cross over
to ~$|I_z|/j \sim j^{-1/2} \ll 1$, which makes the
echo practically zero because it is on the level
of the statistical noise.

The above result indicates that, in order
to revive the quantum phase, which exhibits
the quantum phase diffusion \cite{PD}, it is enough
to exchange {\it coherently} the number of bosons
~$\approx \beta j \sim \sqrt{N}$~ between the components.

In the next section, we will consider
the effect of finite temperature on the echo.

\subsection{Echo at $T\neq 0$}

As discussed in Sec.IIIC, the condensate operators
acting on the eigenstates do not mix them with other
eigenstates. This is stated by Eq.(\ref{I6}), which
is an extension of the Beliaev's result \cite{BEL}
obtained for ~$T=0$, and which is widely employed
at ~$T\neq 0$~ \cite{STAT}. Nevertheless, a special
attention should be focused on calculating the matrix
elements of the time reversal pulse ~${\rm e}^{i\beta I_y}$~
at finite ~$T$, when the normal component is present.
Indeed, a formal separation of the operator ~$I_y \sim N$~
into the condensate and the normal parts could be expected
to produce  a sort of the Debye-Waller thermal factor
~$\exp (-\beta^2N')$, where ~$N'$~ stands for
the number of particles in the thermal cloud.
This logic, however, does not take into account
the symmetry considerations.

We start our analysis of this problem by noting
that, in the case of the exact ~$su(2)$~ symmetry,
the eigenstates ~$|m,N,M\rangle_{(0)}$~ of the
symmetric Hamiltonian ~$H_{(0)}$~ 
form the ~$su(2)$~ representations
for every
value from the set of the excitation quantum
numbers ~$m$. This set should contain also the value ~$j$~
characterizing the dimension of the corresponding representation.
To emphasize this, we will use the notation ~$|m,j,N,M\rangle$~
for the state ~$|m,N,M\rangle_{(0)}$~ characterized by
some value $j$. 
If in the initial state,
all the bosons belong to one component (as in the JILA
experiment \cite{JILA}), this value is
~$j=N/2$, and it remains unchanged during following
evolution. 
This is so because the Hamiltonian forms
a closed algebra with the generators (\ref{3}).
Accordingly, 
 
\begin{eqnarray}
\displaystyle 
\langle M, N,j, m|{\rm e}^{i\beta I_y}|
m',j, N, M'\rangle=d^{(j)}_{M,M'}(\beta)\delta_{m,m'}
\label{T1}
\end{eqnarray}
\noindent
for the initial set of states characterized by some ~$j$.

The situation is different if the intrinsic symmetry
is broken, so that the Hamiltonian does not form
a closed algebra any more. Then, it is natural to expect
that the value ~$j$~ is not a good quantum number under
this circumstance. Furthermore, it is clear that,
at ~$T>T_c$, this value should relax to ~$j\sim \sqrt{N}$,
even though initially it was ~$j=N/2$. Indeed, 
the square of the total angular momentum operator (\ref{3}) 
is ~${\bf I}^2=I_+I_- + I_z^2 - I_z$. At the point
of equal populations, the means ~$\langle I_z \rangle =0,$
~$\langle I_z^2 \rangle 
\sim N$~ and ~$\langle I_+I_- \rangle \sim N$,
where we have employed Eq.(\ref{3}) and 
have taken into account that
no long range order exists above ~$T_c$. This should
be contrasted with the situation at ~$T=0$, when
practically all bosons occupy the condensate
states, and  ~$\langle I_+I_- \rangle \sim N^2$.

What happens at ~$T\neq 0$~ and ~$T<T_c$? 
The dimension of the dominant representation
is given by the numbers of the condensate atoms. 
Indeed, keeping
in mind the explicit form (\ref{3}), one can write
~$\langle I_+I_- \rangle =\langle \int 
d{\bf x}\Psi^{\dagger}_2({\bf x})\Psi_1({\bf x}) 
\int d{\bf x}'\Psi^{\dagger}_1({\bf x}')\Psi_2({\bf x}')
\rangle= N_{c2}N_{c1}(1 +o(1/N))\approx (N/2)^2$~ 
for equal populations in the uniform case. Furthermore,
the correlator ~$\langle I_+(t)I_-(t)I_+(0)I_-(0)\rangle$~
remains time independent as long as the BEC is present.
Indeed, employing Eq.(\ref{I6}), it is easy to show
that this correlator does not depend on time
in the main ~$\sim N^2$~ limit. Thus, the dimension
of the representation is practically selected by the
initial condition. 

Now let us consider how Eq.(\ref{T1}) changes when the ~$su(2)$
symmetry is broken intrinsically, provided the population
difference ~$M$~ is still conserved (due to the absence
of the exchange between the components). 
We employ Eq.(\ref{T1}) in calculating the matrix element
and find
~$\langle M', N, m'|{\rm e}^{i\beta I_y}|
m, N, M\rangle=\sum_{n,j} d^{(j)}_{M',M}(\beta)
\langle M', N, m'| n,j, N, M'\rangle
 \times\langle M, N,j, n|m,N,M\rangle$. This is equal
to ~$\sum_{n,j} d^{(j)}_{M',M}(\beta)
\widehat{\langle M', N, m'}\widehat{| n,j, N, M'\rangle}
\times\widehat{\langle M, N,j, n}\widehat{|m,N,M\rangle}$~
due to the property of the projected states
that their overlaps are equal to the overlaps of the 
corresponding eigenstates for the same values of ~$M,N$
\footnote{ It is important to note that the
projected eigenstates ~$\widehat{|m,j, N, M\rangle}$~
{\it do not} depend on ~$M$. To continue the parallel
between the projected states and the radial part
of the wave function in a central potential (see Sec.III C),
we note that this radial part does not depend on the
~$z$-projection ~$M$~ of the angular momentum.}.
Finally, due to the orthonormality of the projected states
in the subspace with given $j$,
we find

\begin{eqnarray}
\displaystyle 
\langle M', N, m'|{\rm e}^{i\beta I_y}|
m, N, M\rangle=\sum_j^{N/2}d^{(j)}_{M,M'}(\beta)
\widehat{\langle M',N,m'|}{\cal P}_j\widehat{| m,N,M\rangle},
\label{T5}
\end{eqnarray}
\noindent
where ~${\cal P}_j$~ projects the eigenstates to the
space of functions with angular momentum ~$j$, and
the lower limit is either 0 or 1/2, depending on whether
~$N$~ is even or odd, respectively. 
Eq.(\ref{T5}) is a generalization of Eq.(\ref{T1}). 
This equation
gives the representation of the matrix elements 
of the pulse operator in terms of the universal
values ~$ d^{(j)}_{M,M'}(\beta)$~ \cite{LAN}
and the overlaps of the exact projected
eigenstates.
Employing Eq.(\ref{T5})
as well as the orthonormality of the projected
eigenstates for the same ~$M$, we find the 
following representation for Eq.(\ref{90})

\begin{eqnarray}
\displaystyle 
\langle I_z\rangle = \sum_{j_1,j_2,j_3}\sum_{M_{1,2,3,4}}
d^{(j)}_{j,M_1}({\pi \over 2}) d^{(j_1)}_{M_1,M_2}(-\beta)
(I_x)^{(j_2)}_{M_2,M_3} d^{(j_3)}_{M_3,M_4}(\beta) 
&d&^{(j)}_{M_4,j}(-{\pi \over 2})\times
\nonumber
\\
&G&(M_1,M_2,M_3,M_4, t)
\label{T6}
\end{eqnarray}
\noindent
where ~$j=N/2$; ~$(I_x)^{(j_2)}_{M_2,M_3}$~ stands
for the matrix elements of the operator ~$I_x$~ calculated
in the space of the angular momentum ~$j_2$; and
 
\begin{eqnarray}
\displaystyle
&G&(M_1,M_2,M_3,M_4, t)=
\nonumber
\\
\displaystyle
&\sum_m& P_m(N)
\widehat{\langle j,N,m|}{\rm e}^{i{\cal T} H(M_1)}
{\cal P}_{j_1}
{\rm e}^{i(t-{\cal T})H(M_2)} {\cal P}_{j_2}
{\rm e}^{-i(t-{\cal T})H(M_3)} {\cal P}_{j_3}
{\rm e}^{-i{\cal T} H(M_4)}
\widehat{|m,N,j\rangle}\,\, ,
\label{T7}
\end{eqnarray}
\noindent
where we have chosen the initial state
characterized by ~$j=N/2$ (initial populations
$N_2=N,\,\, N_1=0$ \cite{JILA}) and have
suppressed the parameter ~$N$~ 
in the notation ~$H(N,M)$~ of the
projected Hamiltonian; ~$P_m(N)$~ denotes the canonical
normalized thermal distribution  of the
initial states (characterized by ~$M=j,\, j=N/2$).

We note that Eqs.(\ref{T6}), (\ref{T7}) are exact.
They, however, can conveniently be employed only if
the BEC is present. As discussed above, in the case
of the dominant population of the BEC state, the
dimension ~$2j+1$~ of the ~$su(2)$~ representation is
dictated by the initial condition ~$N_2=N,\,\, N_1=0$. 
Accordingly, one can choose ~$j_1=j_2=j_3=j=N/2$ (that is,
omit the external summation
in Eq.(\ref{T6})),
and set
~${\cal P}_j=1$~ in Eq.(\ref{T7}). 
At ~$T=0$, Eq.(\ref{T6}) naturally
transforms into Eq.(\ref{J4}), if one chooses the only
term
~$m=0$~, that is the ground state, in the sum (\ref{T7}).

In what follows, we will only estimate the factor
(\ref{T7}) for the first
echo within the cumulant expansion, which was
employed in Sec.IIIC and is outlined in
Appendix B. We consider temperatures low enough,
so that the fluctuations of the condensate
populations \cite{FLUK} can be ignored\footnote{
Otherwise, the complete form (\ref{T6})
with the summation over ~$j_{1,2,3}$~ should be analyzed}.
  
The suppression of the echo at ~$t=2{\cal T}$~ is
described by the factor (\ref{T7}), where we set
~$t=2{\cal T}$. 
Employing Eq.(\ref{B8}) as
shown in Appendix B, we find  

\begin{eqnarray}
\displaystyle 
&\langle& I_z(2{\cal T})\rangle
\approx
 j\sum_{M,M'}
d^{(j)}_{j,M}(\pi/2) d^{(j)}_{M+1,j}(-\pi /2)
J_{M'-M}(\beta j)
J_{M'-M-2}(\beta j)\times 
\nonumber
\\
\displaystyle
&\sum_{m}& \tilde{P}_{m}(M)
{\rm e}^{i{\cal T}\{ E_m(M-1)- E_m(M)+ E_m(M')- E_m(M'-1)
- [2(M-M'-{1\over 2})^2 +{3\over 2}]t^{-1}_{OC}\}},
\label{T8}
\end{eqnarray}
\noindent
where we have taken into account that the effective
values of ~$M,\,\, M'$~ are much less than ~$j=N/2$;
the definition of the effective distribution function
~$\tilde{P}_{m}(M)$~ created by the first pulse
is given in Appendix B, Eq.(\ref{B3}).
We have approximated
the matrices of finite rotations by the Bessel
functions similarly to how it was done
in Eq.(\ref{J4}). The matrix ~$I_x$~ has also been
replaced by ~$j/2$~ in this space, and,
accordingly, the terms ~$M_3=M_2\pm 1$~
have been selected.

We note that Eq.(\ref{J4}) (with ~$t=2{\cal T}$~)
is, in fact,
the limiting case ~$t_{OC} \to \infty$~ 
of Eq.(\ref{T8}). Thus, we have actually assumed
that the limit ~$t_{OC} \to \infty$~ corresponds
to ~$T=0$, and, conversely, ~$t_{OC}$~ becomes
small enough to include its effect at
high ~$T$. Here we will not calculate 
a specific expression for ~$t_{OC}$ as a function
of ~$T$.
 
In order to obtain a closed
form expression for 
Eq.(\ref{T8}), we employ the smallness
~$|M/N|\ll 1,\,\, |M'/N|\ll 1$, and, accordingly,
expand the eigenenergies as

\begin{eqnarray}
\displaystyle 
E_m(M)&=&E_m(0)+
\epsilon_mM+{1\over 2}b_{2m}M^2 +o(1/N^2),
\label{T91}
\\
 \epsilon_m&=&{\partial E_m(M)\over
\partial M}_{|M=0},\,\,\,b_{2m}=
{\partial^2 E_m(M)\over
\partial M^2}_{|M=0} 
\label{T9}
\end{eqnarray}
\noindent
 around the point ~$M=0$~ of equal populations.
Then, Eq.(\ref{T8})
becomes

\begin{eqnarray}
\displaystyle 
\langle I_z(2{\cal T})\rangle
\approx
 j\sum_{K}
J_K(\beta j)
J_{K-2}(\beta j) 
{\rm e}^{i{\cal T}\{\overline{b}_2K
- [2(K-{1\over 2})^2 +{3\over 2}]
t^{-1}_{OC}\}},
\label{T10}
\end{eqnarray}
\noindent
where we have ignored the fluctuations
of ~$b_{2m}\sim 1/N$~ and wrote
~$
\sum_{m} \tilde{P}_{m}(M)
\exp[i{\cal T} (M' -M)b_{2m}]=
\exp[i{\cal T} (M' -M)\overline{b}_2]$, with
~$\overline{b}_2=
\sum_{m} \tilde{P}_{m}(0)
b_{2m}$.
It is worth noting, that, in the limit
~$t_{OC}=\infty$, Eq.(\ref{T10}) yields
the first echo magnitude (\ref{J7}) (where
the value ~$b_2=b_{2m=0}$~ is
replaced by ~$\overline{b}_2$). 
We have also taken into account
that ~$\sum_M
d^{(j)}_{j,M}(\pi/2) d^{(j)}_{M+1,j}(\pi /2)
=1$~ with a very good accuracy in the limit
~$j\gg 1$.

The exponential factor in Eq.(\ref{T10}), 
containing ~$t_{OC}^{-1}$, produces some suppression
of the echo strength.
As can be verified numerically, the sum (\ref{T10})
at its maximum decreases from the value about
0.45, which corresponds to ~$t_{OC}=\infty$, to
the value about 0.06, corresponding  to 
~$t_{OC}\approx 2{\cal T}$. A change
is also exhibited by the value of ~$\beta j$~ at the
maximum of the echo. It shifts from the value given by
Eq.(\ref{J8}) ~$\beta \approx 1/\sqrt{j}$~
(in the limit ~$ t_{OC}=\infty$)
to ~$\beta \approx 4.5/j$ for ~$ t_{OC}\approx 2{\cal T}$. 
This can be understood as follows,
as  ~$t_{OC}\to 2{\cal T}$, only few Bessel functions
depending on the product ~$\beta j$
contribute to the sum (\ref{T10}),
and ~$\beta j \approx 4.5$ corresponds to the maximum
of this contribution. As we discussed in Sec.IIIC,
~$t_{OC} \sim N$~ as long as the population of the 
condensate is comparable to ~$N$. Thus, practically,
~$t_{OC} \gg {\cal T} \approx \tau_d \sim \sqrt{N}$.
This implies that the thermal effects 
do not suppress the echo substantially, if compared
with the zero ~$T$~ limit. 
It should, however, be noted that
at temperatures high enough, the fluctuations of the 
condensate population \cite{FLUK} would lead to
the necessity of including the summation
over ~$j_{1,2,3}$~ in Eq.(\ref{T6}).  This problem
will be analyzed elsewhere.

We also not that external factors, which introduce
irreversible decoherence, may suppress the echo significantly.
If a typical time of the loss of the coherence due to
these factors is ~$\tau_{irr}$, as given by, e.g., Eq.(\ref{D10}),
the first echo intensity will acquire the exponential
factor ~$\sim {\rm e}^{- 2{\cal T}/\tau_{irr}}$, and, accordingly,
no echo will be practically 
observed in the situation ~${\cal T} > \tau_{irr}$.

\section{Summary}

We have shown that the intrinsic symmetry
in the multi-component atomic mixtures 
has critical impact on the decoherence of
the Ramsey fringes. This symmetry should be broken
in one way or another in order to induce damping.
The extrinsic factors, such as, e.g., the interaction with 
the background gas, can produce 
the decoherence of the
inter-component correlator. This decoherence,
however,
is not faster than the rates of the heating and of the 
induced losses.  

We have shown that, in the two-component
BEC, the intrinsic decoherence is 
a result of the ensemble fluctuations of the 
relative chemical potential. In other words,
the evolution of the condensate operators occurs
as though the normal component is not affected
by it.
Therefore, this evolution can be considered
reversible.

We have analyzed the atomic echo effect
in the Ramsey spectroscopy of the
two-component BEC. The strength of the
echo is obtained in the case of the
phase diffusion at zero and finite
temperatures.
We have found that the echo
survives finite temperatures, as long as
the population of the condensate  state
remains dominant.

We also point out that
the echo effect depends essentially on
the nature of the many-body correlations,
and as well as on the 
deviations from the intrinsic ~$su(2)$~ symmetry.
Accordingly, its measurement 
would provide valuable information on the nature
of the many body correlations in the trapped multi component BEC.  
Because of this, we suggest that the atomic echo effect
should be studied experimentally.

\acknowledgements
We acknowledge stimulating discussions with
E. A. Cornell, D. S. Jin, Y. M. Kagan, 
A. Ruckenstein, and D. Schmeltzer. 
We are grateful to L.P. Pitaevskii for
useful discussions of the results, 
as well as to M. D. Girardeau,
L. You and E. Zaremba
for interest to the work. 
 This work was partially
supported by the CUNY Research Program.

\appendix
\section{Equation for OPDM in the presence
of white noise}

Furutzu-Novikov \cite{NOV} theorem states that 

\begin{eqnarray}
\displaystyle 
\langle \xi(x,t) Z(x',t')\rangle =
\int \, dy \int \, d\tau 
\langle \xi(x,t) \xi (y, \tau)\rangle
\langle {\delta Z(x',t')\over \delta \xi(y, \tau)}\rangle,
\label{A1}
\end{eqnarray}
\noindent
where the averaging is performed over the
gaussian noise ~$\xi (x,t)$; ~$Z(x,t)$~ is an
arbitrary smooth functional of ~$\xi$.
This relation can be verified by expanding
~$Z(x,t)$~ in the functional powers of 
~$\xi$~ and by comparing the l.h.s. and the r.h.s.
 term by term. 

We apply (\ref{A1}) to the Heisenberg
equations, following from Eq.(\ref{D4})
and written for the operators
~$\hat{\rho}_{ij}
({\bf x},{\bf x}', t)=\Psi^{\dagger}_i
({\bf x},t) \Psi_j({\bf x}', t)$.

\begin{eqnarray}
\displaystyle
i\hbar\dot{\hat{\rho}}_{ij}({\bf x},{\bf x}', t)=
[-H_{0{\bf x}} -\xi_i({\bf x},t) + H_{0{\bf x}'}
+ \xi_j( {\bf x}',t)] \hat{\rho}_{ij}
({\bf x},{\bf x}', t) ,\,\,\, \xi_1=-\xi_2=\xi,
\label{A2}
\end{eqnarray}
\noindent
where no two-body interaction is taken into account.
We average Eq.(\ref{A2}) over the initial condition and
the noise, and obtain
for the OPDM ~$\rho_{ij}
({\bf x},{\bf x}', t)=\langle \hat{\rho}_{ij}
({\bf x},{\bf x}', t)\rangle$:

\begin{eqnarray}
\displaystyle
i\hbar \dot{\rho}_{ij}({\bf x},{\bf x}', t)=
[-H_{0{\bf x}} + H_{0{\bf x}'}]
\rho_{ij}({\bf x},{\bf x}', t)
+ \langle [-\xi_i({\bf x},t) + \xi_j({\bf x},t)]
\rho_{ij}({\bf x},{\bf x}', t)\rangle.
\label{A3}
\end{eqnarray}
\noindent
Then, we represent the mean ~$
\langle [-\xi_i({\bf x},t) + \xi_j({\bf x},t)]
\rho_{ij}({\bf x},{\bf x}', t)\rangle$~ in accordance
with Eq.(\ref{A1}), and write down the equation
for the functional derivative as

\begin{eqnarray}
\displaystyle
i\hbar {\partial \over \partial t}
{\delta \rho_{ij}({\bf x},{\bf x}', t)\over
\delta\xi ({\bf y}, \tau )}&=&
[-H_{0{\bf x}} + H_{0{\bf x}'}]
{\delta \rho_{ij}({\bf x},{\bf x}', t)\over
\delta \xi ({\bf y}, \tau )}
+ \langle [-\xi_i({\bf x},t) + \xi_j({\bf x},t)]
{\delta \rho_{ij}({\bf x},{\bf x}', t)\over
\delta \xi ({\bf y}, \tau )}\rangle 
\nonumber
\\
\phantom{XXX}
\label{A4}
\\
&+& [- \delta( {\bf x} -{\bf y}) +
\delta( {\bf x}' -{\bf y})]
\langle \rho_{ij}({\bf x},{\bf x}', t)\rangle
\delta (t-\tau).
\nonumber
\end{eqnarray}
\noindent

We take into account the causality, so that
~$ \delta \rho_{ij}({\bf x},{\bf x}', t)/
\delta \xi ({\bf y}, \tau )=0$~ for ~$t<\tau$.
Then, we find

\begin{eqnarray}
\displaystyle
{\delta \rho_{ij}({\bf x},{\bf x}', t)\over
\delta \xi ({\bf y}, \tau =t-0 )}= -
{i\over \hbar} [- \delta( {\bf x} -{\bf y}) +
\delta( {\bf x}' -{\bf y})]
\langle \rho_{ij}({\bf x},{\bf x}', t)\rangle
\label{A5}
\end{eqnarray}
\noindent
from Eq.(\ref{A4}). We also take advantage
of the white noise structure (\ref{D3}). 
Finally, substituting Eq.(\ref{D3}) and Eq.(\ref{A5})
into Eq.(\ref{A1}), we obtain Eqs.(\ref{D61}),
(\ref{D63}).

\section{Thermal suppression of the echo}

To simplify the notations,
we will be suppressing ~$N$ -- the total number
of bosons. We take
~$j_{1,2,3}=j$, and represent the factor (\ref{T7}) at 
~$t=2\cal T$~, and for ~$M_3=M_2 - 1$~
(the term with ~$M_3=M_2+1 $~ is just the
complex conjugate) as

\begin{eqnarray}
\displaystyle 
G(M_1,M_2,M_2-1,M_4, 2{\cal T})=
\sum_{m'} &P_{m'}&
\widehat{\langle j,m'|}\widehat{m,M_4\rangle}
\widehat{\langle M_4,n|}\widehat{m',j\rangle}\times
\nonumber
\\
\widehat{\langle m,N,M_4|}{\rm e}^{i{\cal T} H(M_1)}
&{\rm e}&^{i{\cal T}H(M_2)}{\rm e}^{-i{\cal T}H(M_2-1)}
{\rm e}^{-i{\cal T} H(M_4)}
\widehat{|n,M_4\rangle}\,\, ,
\label{B1}
\end{eqnarray}
\noindent
and select the diagonal terms (with respect to the
excitation label) only ~$m=n$. In fact, the evolution
of the off-diagonal terms is controlled by 
the excitations, and, therefore, they
decay rapidly on the chosen time scale ~$\tau_d$. Then,
Eq.(\ref{B1}) acquires the form: 

\begin{eqnarray}
\displaystyle 
G(M_1,M_2,M_2-1,M_4, 2{\cal T})&=&
\nonumber
\\
\sum_{m} \tilde{P}_{m}(M_4)
\widehat{\langle m,M_4|}&{\rm e}&^{i{\cal T} H(M_1)}
{\rm e}^{i{\cal T}H(M_2)}{\rm e}^{-i{\cal T}H(M_2-1)}
{\rm e}^{-i{\cal T} H(M_4)}
\widehat{|m,M_4\rangle}\,\, ,
\label{B2}
\\
\tilde{P}_{m}(M_4)&=&\sum_{m'}P_{m'}
\widehat{\langle j,m'|}\widehat{m,M_4\rangle}
\widehat{\langle M_4,m|}\widehat{m',j\rangle}.
\label{B3}
\end{eqnarray}
\noindent
The quantity ~$\tilde{P}_{m}(M_4)$~ can be considered
as an effective distribution function created by the first pulse for 
the given population difference ~$2M_4$. As it can
be seen, it satisfies the normalization condition
~$\sum_m\tilde{P}_{m}(M_4)=1$. 

As seen from Eqs.(\ref{J4}), (\ref{J6}), 
the first echo is dominated
by the term ~$M_4=M_1+1$. Accordingly, in what follows we will
consider the term ~$ 
G(M-1,M',M'-1,M, 2{\cal T})$~
only. We apply the cumulant expansion
in order to evaluate it.
Thus,

\begin{eqnarray}
\displaystyle 
G(M-1,M',M'-1,M, 2{\cal T})
&=&
\sum_{m} \tilde{P}_{m}(M)
\widehat{\langle M,m|}
{\rm e}^{-i{\cal T}H(M)}{\rm e}^{i{\cal T}H(M-1)}
\times
\nonumber
\\
&{\rm e}&^{i{\cal T}H(M')}{\rm e}^{-i{\cal T}H(M'-1)}
\widehat{|m,M\rangle}=G_{1c}G_{2c}.
\label{B4}
\end{eqnarray}
\noindent
The first cumulant 

\begin{eqnarray}
\displaystyle
G_{1c}=\sum_{m} \tilde{P}_{m}(M)
{\rm e}^{i{\cal T}[ E_m(M-1)- E_m(M)+ E_m(M')- E_m(M'-1)]}
\,\, (B41)
\label{B41}
\end{eqnarray}
\noindent
corresponds to the assumption
~$\widehat{\langle M,m }
\widehat{|m',M'\rangle} =
\delta_{m,m'}$~ for any value of the difference
~$M-M'$~ akin to Eq.(\ref{I6}). This, however,
is not completely the case. In fact, for large values
of ~$M-M'$, a sort of the Debye-Waller factor
enters this product. The second cumulant
~$G_{2c}$~ takes care of this effect. 

We employ the identity 

\begin{eqnarray}
\displaystyle 
{\rm e}^{-iH(M'){\cal T}}&=&
{\rm e}^{-iH(M) {\cal T}}{\rm Texp} [-i(M'-M)
\int^{\cal T}_0\, dt' H'(t')],
\label{B5}
\\
H'(t)&=&{\rm e}^{iH(M)t}\{{\partial H(M) 
\over \partial M} + o(1/N)\}{\rm e}^{-iH(M)t},
\label{B6}
\end{eqnarray}
\noindent
where we have followed the same reasoning which
lead us to Eq.(\ref{I2}) -- keeping the first
term of the expansion of the projected Hamiltonian.

We are interested in evaluating the effect of
suppression of the diagonal matrix
elements (with different ~$M$~ and the same ~$m$)
on the time evolution.
Then, we write
~$\widehat{\langle M,m|}
{\rm e}^{-i{\cal T}H(M)}{\rm e}^{i{\cal T}H(M-1)}
{\rm e}^{i{\cal T}H(M')}{\rm e}^{-i{\cal T}H(M'-1)}
\widehat{|m,M\rangle}\approx 
\widehat{\langle M,m|}
{\rm e}^{i{\cal T}H(M-1)}\widehat{|m,M\rangle}
\widehat{\langle M,m|}
{\rm e}^{-i{\cal T}H(M)}\widehat{|m,M\rangle}
\widehat{\langle M,m|}
{\rm e}^{i{\cal T}H(M')}\widehat{|m,M\rangle}
\widehat{\langle M,m|}
{\rm e}^{i{\cal T}H(M'-1)}\widehat{|m,M\rangle}$, which
after employing Eqs.(\ref{B5}),(\ref{B6})
for the each mean, and substituting back
to Eq.(\ref{B4}), yields the first (\ref{B41})
and second cumulant

\begin{eqnarray}
\displaystyle
G_{2c}= \exp \{- [2(M-M'-{1\over 2})^2 +{3\over 2}&]&\times
\nonumber
\\
\int^{\cal T}_0\, dt'\int^{t'}_0\, dt
&\sum_{m}& \tilde{P}_{m}(M)\widehat{\langle M,N,m|}
\delta H'(t) \delta H'(0)
 \widehat{|m,N,M \rangle} \},
\label{B7}
\end{eqnarray}
\noindent
where ~$\delta H'(t)$~ is defined below Eq.(\ref{I3}).
We employ the definition of ~$t_{OC}$ in Eq.(\ref{I5}), and
rewrite Eq.(\ref{B4}) as

\begin{eqnarray}
\displaystyle 
&G&(M-1,M',M'-1,M, 2{\cal T})=
\nonumber
\\
&\sum_{m}& \tilde{P}_{m}(M)
{\rm e}^{i{\cal T}\{ E_m(M-1)- E_m(M)+ E_m(M')- E_m(M'-1))
- [2(M-M'-{1\over 2})^2 +{3\over 2}]
t^{-1}_{OC}\}},
\label{B8}
\end{eqnarray}
\noindent
which is valid in the 
limit ~$\tau_n \leq {\cal T} \ll t_{OC}$. 
This expression is to be employed
in Eq.(\ref{T6}).

\begin{figure}[hbt]
\epsfxsize=\columnwidth\epsfbox{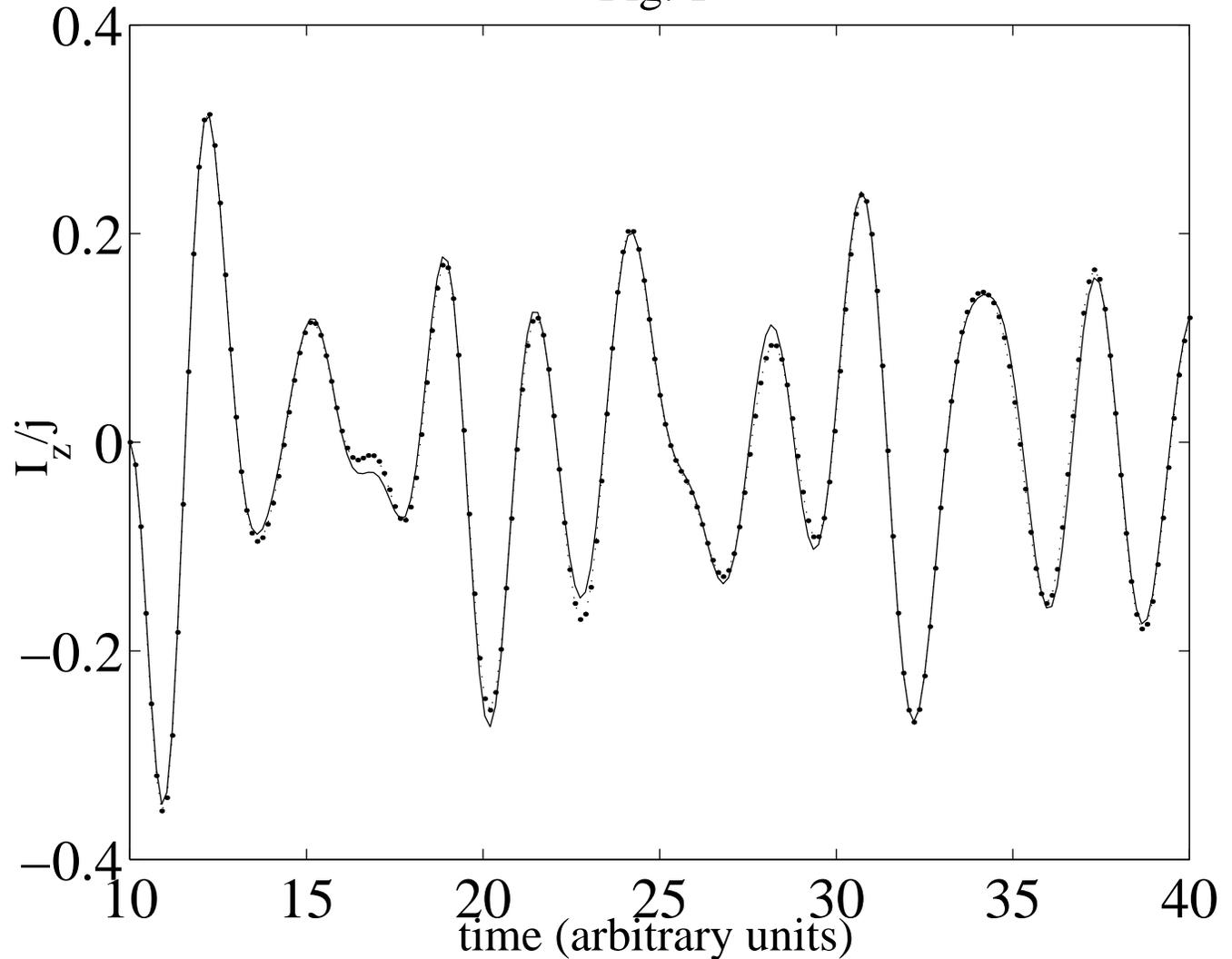}
\caption{ 
Comparison of the results of the exact calculations
of Eq.(\ref{90}) with that obtained within the
quantum phase method. Solid line corresponds
to the quantum phase method. Dotted line --
the exact calculations. Parameters chosen:
$j=45$ (total number of particles $N=2j=90$),
$\beta =0.22,\,\, {\cal T}=10,\,\, b_2=0.05,\,\,
\epsilon_0=2$.
}
\label{fig1}
\end{figure}

\begin{figure}[hbt]
\epsfxsize=\columnwidth\epsfbox{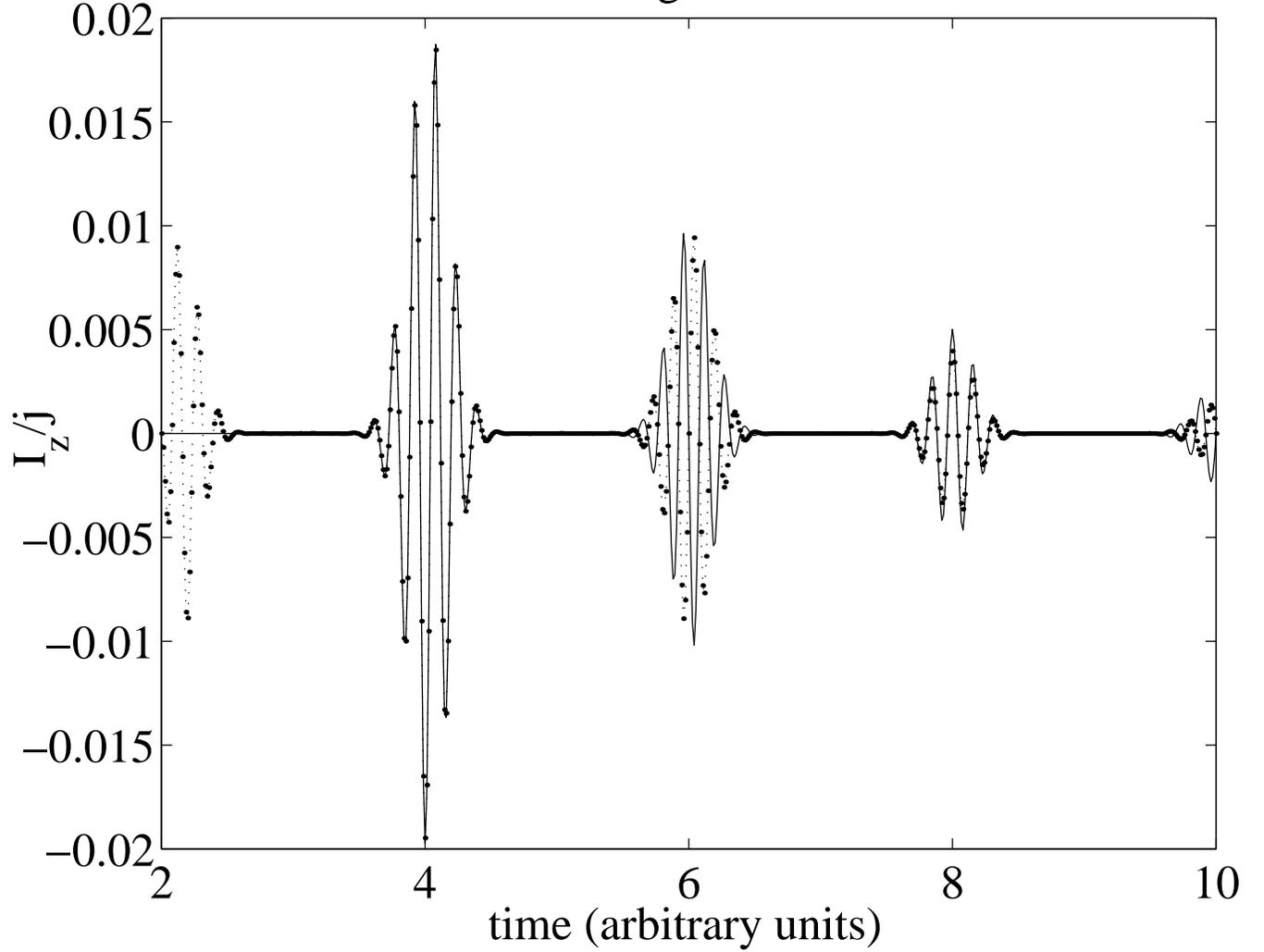}
\caption{ 
The
echo effect in a confined
two-component
Bose-Einstein condensate
at ~$T=0$. Vertical
axis is labeled by the relative magnitude
of the population difference after the read-out
pulse: $I_z$ is the half of the population
difference; $j$ stands for the half of
the total number of bosons.
Solid line -- the analytical solution (\ref{J6});
Dotted line -- the numerical solution
of Eq.(\ref{J4}). Parameters chosen:
~$j=2000$ (total number of particles $N=2j=4000$),
~${\cal T}=2,\,\, b_2=0.1,\,\,
\epsilon_0=40$;
~$\beta =0.0005,\,\, \beta j=1$}
\end{figure}

\begin{figure}[hbt]
\epsfxsize=\columnwidth\epsfbox{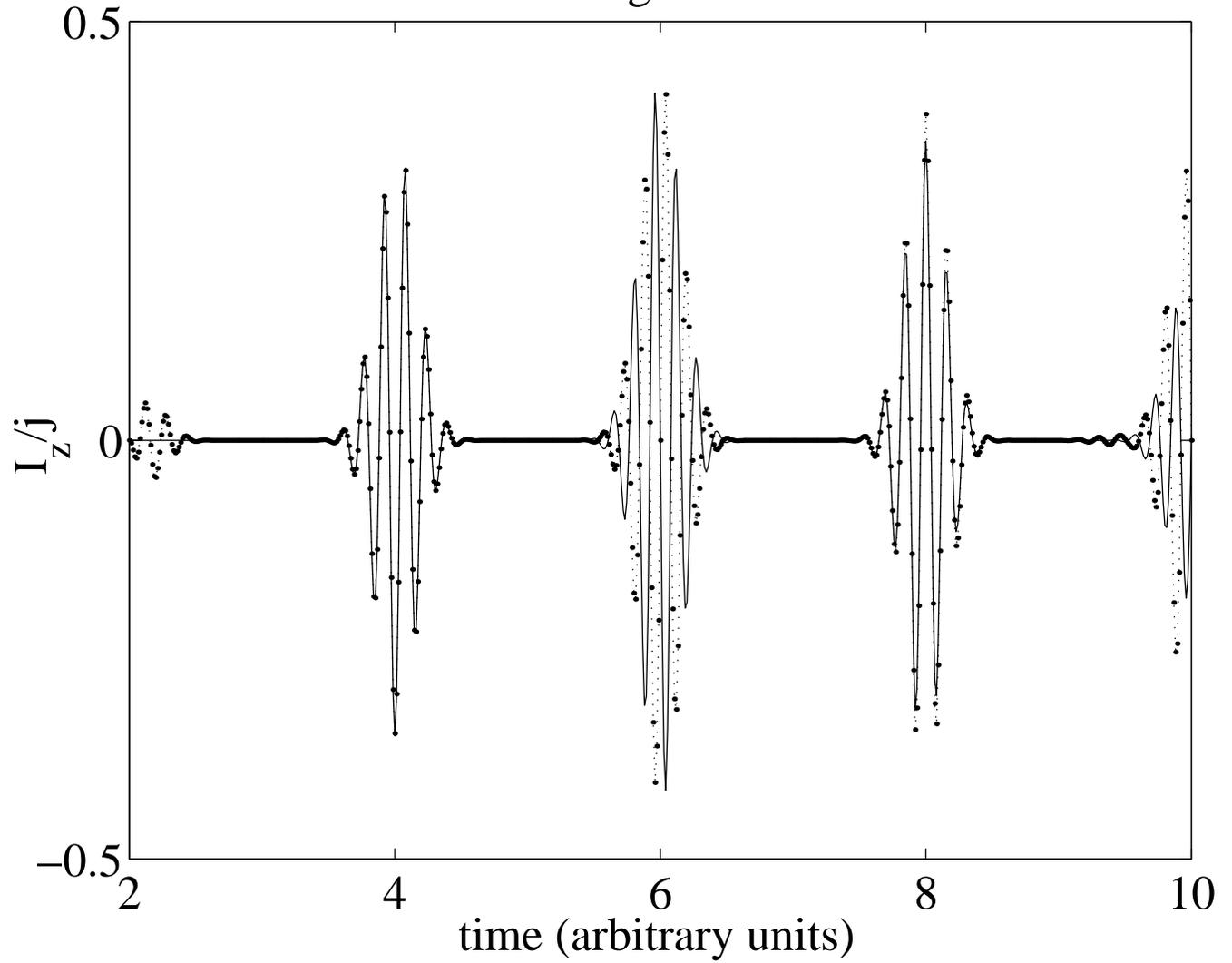}
\caption{ 
Same as Fig2a except the values
~$\beta =0.0025,\,\, \beta j=5$
}
\end{figure}

\end{document}